\documentclass[sigconf]{acmart}
\usepackage{booktabs}
\usepackage{colortbl}
\usepackage{multirow}
\usepackage{tikz}
\usetikzlibrary{positioning,arrows.meta,calc,decorations.pathreplacing}
\usepackage{algorithm}
\usepackage{algpseudocode}
\usepackage{cleveref}
\usepackage{placeins}
\usepackage{enumitem}
\setcopyright{none}
\acmConference[Submitted for review]{}{2027}{}
\renewcommand\footnotetextcopyrightpermission[1]{}

\newtheorem{proposition}{Proposition}

\newtheorem{lemma}{Lemma}

\newcommand{\bestParams}{52}

\newcommand{\burstGap}{20}
\newcommand{\burstMaePers}{22.6}
\newcommand{\burstMaePh}{14.1}
\newcommand{\burstMaeSes}{17.0}
\newcommand{\burstN}{137}
\newcommand{\burstNow}{46}

\newcommand{\burstObsTwo}{27}

\newcommand{\burstPredTwo}{30}

\newcommand{\caseTeam}{8}

\newcommand{\deMaxTrain}{16}
\newcommand{\deMinTrain}{2}

\newcommand{\dyUsefulHolmSec}{96}

\newcommand{\ewAucAtt}{0.76}
\newcommand{\ewAucConv}{0.85}

\newcommand{\ewAucProx}{0.66}

\newcommand{\ewBestOtherProx}{0.68}

\newcommand{\ewDetConv}{74}

\newcommand{\ewFar}{20}

\newcommand{\ewWindowSec}{64}
\newcommand{\fracSilentMember}{44}

\newcommand{\hierBalDerived}{2.86}
\newcommand{\hierBalExplicit}{0.77}

\newcommand{\hierBalRatio}{3.7}

\newcommand{\ivEffMpc}{1.51}
\newcommand{\ivEffPeriodic}{0.70}

\newcommand{\ivEffRatioMpcPeriodic}{2.2}
\newcommand{\ivEffRatioMpcReactive}{1.3}
\newcommand{\ivEffReactive}{1.17}

\newcommand{\ivMpcWins}{8}
\newcommand{\ivMpcWinsPh}{6}
\newcommand{\ivMpcWinsVar}{2}

\newcommand{\jMax}{0.05}
\newcommand{\longMax}{0.79}
\newcommand{\maxWindows}{74}
\newcommand{\minWindows}{21}
\newcommand{\nGroups}{17}

\newcommand{\nWindows}{688}
\newcommand{\overlapRatio}{1.08}
\newcommand{\passiveFrac}{100}

\newcommand{\rbCleanMlp}{1.16}

\newcommand{\rbCleanPh}{1.01}

\newcommand{\rbCleanVar}{1.12}

\newcommand{\rbGaussTwoPh}{1.14}

\newcommand{\rbGaussTwoShiftPh}{2.97}

\newcommand{\rbGaussTwoShiftVar}{1.29}

\newcommand{\rbGaussTwoVar}{1.38}
\newcommand{\rhoAttCov}{-0.46}
\newcommand{\rhoAttEps}{-0.44}
\newcommand{\rhoAttStab}{-0.47}
\newcommand{\rhoContraction}{0.814}
\newcommand{\rhoGainTwo}{3.0}
\newcommand{\rhoHProgress}{-0.08}
\newcommand{\rhoHalfLifeSec}{54}

\newcommand{\sensSettings}{12}
\newcommand{\symDyChg}{+0.1}
\newcommand{\symDyP}{0.963}
\newcommand{\symTmChg}{-0.0}
\newcommand{\symTmP}{0.890}
\newcommand{\tauConsAtt}{17.0}
\newcommand{\tauConsConv}{25.9}
\newcommand{\tauConsProx}{27.6}

\newcommand{\tmUsefulHolmSec}{80}

\newcommand{\topDyadBest}{30}
\newcommand{\topDyadPers}{31}

\newcommand{\wPopConv}{0.37}
\newcommand{\wPopProx}{0.50}

\newcommand{\wSessionAtt}{0.81}

\newcommand{\wSessionProx}{0.41}

\newcommand{\wTeamPopSpeakingBalance}{0.56}

\newcommand{\wTeamRecentSpeakingBalance}{0.43}

\newcommand{\wTeamSessionSpeakingBalance}{0.02}

\newcommand{\clDiam}{4.77}
\newcommand{\clPort}{1.01}
\newcommand{\clKappa}{1.61}
\newcommand{\clCertShort}{5.2}
\newcommand{\clCertSession}{6.8}
\newcommand{\clCertLong}{8.2}
\newcommand{\clCertEvery}{12.9}
\newcommand{\clSteady}{138}
\newcommand{\clFormulaErr}{1}
\newcommand{\clSimAtt}{5.9}
\newcommand{\clSimSpk}{2.0}
\newcommand{\clSimProx}{2.5}
\newcommand{\clSimAttLong}{12.6}
\newcommand{\clWpopAtt}{0.012}

\newcommand{\memObsc}{24}
\newcommand{\memPredc}{30}

\newcommand{\memSurr}{-1}

\newcommand{\enOrigins}{122}
\newcommand{\enObsThree}{0.65}
\newcommand{\enSurrThree}{0.70}
\newcommand{\enPredThree}{0.06}

\newcommand{\memSpkTwo}{24}
\newcommand{\memAttTwo}{10}
\newcommand{\memProxTwo}{29}

\newcommand{\tauConsBal}{12}
\newcommand{\tauExplBal}{22}
\newcommand{\sesOnlyDyP}{0.035}
\newcommand{\sesOnlyTmP}{0.020}
\newcommand{\relStaticDyS}{+11.3}
\newcommand{\relStaticDySP}{2e-5}

\newcommand{\relStaticTmS}{+8.5}
\newcommand{\relStaticTmSP}{5e-4}

\newcommand{\relPersDyL}{+49.2}

\newcommand{\relPersTmL}{+43.5}

\newcommand{\relEmaDyS}{+23.1}

\newcommand{\relRidgeAugDyL}{+1.5}
\newcommand{\relRidgeAugDyLP}{0.306}

\newcommand{\relRidgeAugTmL}{+1.4}
\newcommand{\relRidgeAugTmLP}{0.487}

\newcommand{\relScalarDyS}{+1.7}
\newcommand{\relScalarDySP}{0.006}

\newcommand{\relScalarTmS}{+0.0}

\newcommand{\relFreeDyS}{+1.8}

\newcommand{\relFreeTmS}{+4.4}

\newcommand{\hzAnchorOneTwoEight}{1.10}
\newcommand{\hzRefOneTwoEight}{1.32}
\newcommand{\bowlLo}{0.24}
\newcommand{\bowlHi}{0.54}
\newcommand{\bowlPop}{0.44}
\newcommand{\bowlTeam}{8}

\newcommand{\arAllReal}{24}
\newcommand{\arAllNull}{37}
\newcommand{\arAllP}{1.5e-5}
\newcommand{\arAllWins}{17}
\newcommand{\arTeamReal}{17}
\newcommand{\arTeamNull}{29}
\newcommand{\arTeamP}{0.008}
\newcommand{\arTeamWins}{13}

\newcommand{\rtFits}{170}
\newcommand{\rtFreeMax}{0.92}
\newcommand{\rtRhoMax}{0.998}
\newcommand{\flZeroKappa}{1.76}
\newcommand{\flZeroDy}{1.082}
\newcommand{\flZeroTm}{1.002}
\newcommand{\flEightKappa}{0.96}
\newcommand{\flEightBound}{20.6}
\newcommand{\flEightDy}{1.201}
\newcommand{\flEightTm}{1.115}
\newcommand{\flNineKappa}{0.89}
\newcommand{\flNineBound}{7.7}
\newcommand{\flNineDy}{1.230}
\newcommand{\flNineTm}{1.143}

\newcommand{\phOnce}{15}
\newcommand{\phFirstSec}{304}
\newcommand{\phFirstMin}{5}
\newcommand{\phFirstLow}{288}
\newcommand{\phFirstHigh}{320}
\newcommand{\phSoon}{11}
\newcommand{\phAny}{26}
\newcommand{\phAgree}{82}
\newcommand{\phCorrWin}{$-0.44$}
\newcommand{\phCorrRest}{$-0.83$}

\newcommand{\pH}{pH}
\newcommand{\chiFit}{0.08}
\newcommand{\rhoGainOne}{5.4}
\newcommand{\model}{ANCHOR}
\definecolor{ours}{HTML}{E4ECF7}
\newcolumntype{O}{>{\columncolor{ours}}c}
\newcommand{\ann}[2]{\underbrace{#1}_{\text{\scriptsize #2}}}


\begin{document}

\title[ANCHOR]{ANCHOR: A Port-Hamiltonian Model for Forecasting Team Dynamics in Human-in-the-Loop Systems}
%\title[ANCHOR]{ANCHOR: Port-Hamiltonian Hypothesis of Team Dynamics for Human-in-the-Loop Systems.}
\author{Diana Romero}
   \email{dgromer1@uci.edu@uci.edu}
%   \orcid{0000-0002-5579-1255}
   \affiliation{%
   \institution{University of California, Irvine}
   \city{Irvine}
   \state{CA}
   \country{USA}
 }
\author{Salma Elmalaki}
   \email{salma.elmalaki@uci.edu}
%   \orcid{0000-0002-5579-1255}
   \affiliation{%
   \institution{University of California, Irvine}
   \city{Irvine}
   \state{CA}
   \country{USA}
 }
%\affiliation{\institution{University of California, Irvine}\country{}}

\begin{abstract}
Cyber-physical-human systems close a loop around people by sensing human behavior, predict it and act on it. Human teams are a natural target, because collaboration breaks down in ways that a timely prompt could prevent. Collaborative mixed reality makes this loop practical, since each headset senses who speaks, where its wearer looks and who stands together, and it can deliver a light prompt. In this loop, the headsets are sensors, the prompts are actuators and the team is the plant. Closing the loop needs a plant model that is small enough to learn from few teams, safe to iterate and readable by a designer. We propose \model{}, which models a team as a ball in an energy bowl. The bottom of the bowl is where the team rests, and it is set by the team's recent behavior, its own history and other teams. Formally, \model{} is a switched port-Hamiltonian system with \bestParams{} parameters, and each part stands for a testable claim from team psychology. Because of this structure, for any fitted parameters, a forecast moves away from rest only when task progress or a prompt pushes it, and prompts of bounded size that are spaced in time move it a bounded distance. On \nGroups{} four-person teams held out from training, a team that drifts from its usual behavior returns most of the way within half a minute. After 32\,s, about \memObsc\% of the deviation remains, and \model{} predicted \memPredc\%. Forecasts are clearly better when the resting point blends its three sources than when it uses one. \model{} matches the accuracy of a black-box baseline with about 40,000 parameters. It anticipates conversation collapse with an AUROC of \ewAucConv. In simulation under assumed prompt effects, it removes \ivEffRatioMpcPeriodic{} times as much breakdown time per prompt as periodic prompting.

\end{abstract}



\maketitle

\section{Introduction}\label{sec:intro}

Many goal-oriented tasks are carried out by teams, from operating rooms and control rooms to design reviews and training. How well a team performs depends not only on its members' skills but also on how they interact, for example on whether everyone gets to speak and whether members attend to the same things~\cite{woolley2010,kozlowski2006}. These patterns change during a session, and a team can drift into a breakdown in which one member dominates or the team loses its shared focus. A timely prompt can correct such a drift, and coaching helps most when it arrives at the right moment~\cite{hackman2005}. A system that senses a team, anticipates its breakdowns and prompts it is therefore a cyber-physical-human system in which people are part of the loop~\cite{schirner2013,bengler2024hcps}.

\begin{figure}[t]
\centering
\begin{tikzpicture}[font=\scriptsize, >={Latex[length=1.4mm]}]
\fill[blue!5] plot[domain=-1.55:1.55, samples=40] (\x, {0.55*\x*\x}) -- cycle;
\draw[thick] plot[domain=-1.6:1.6, samples=40] (\x, {0.55*\x*\x});
\fill[orange!70!black] (1.05, {0.55*1.05*1.05+0.11}) circle (0.11);
\node (now) at (0.1, 1.17) {team now};
\draw[->, gray] (now.east) -- (0.97, 0.84);
\draw[->, thick, blue!60!black] (0.9, 0.55) to[bend left=10] (0.25, 0.09);
\node[align=center, blue!60!black] at (-0.3, 0.66) {regulation $R$\\pulls it back};
\draw[densely dotted] (0, 0) -- (1.32, 0);
\draw[densely dotted] (1.32, 0) -- (1.32, 0.71);
\draw[<->] (1.4, 0) -- node[right, align=left] {energy\\$H$} (1.4, 0.71);
\fill (0, 0) circle (0.04);
\node[below=1pt, align=center] at (0, 0) {resting point $q^\star$, anchored on\\recent behaviour, team history\\and other teams};
\draw[<->, gray] (-1.5, -0.95) -- node[below] {task progress moves the bowl} (1.5, -0.95);
\begin{scope}[xshift=3.45cm, yshift=1.5cm, node distance=1.6mm,
  box/.style={draw, rounded corners=1.5pt, align=center, inner sep=1.8pt, text width=2.05cm, fill=#1}, box/.default=white]
\node[box=gray!8] (s) {MR headsets: speech,\\gaze, pose, task logs};
\node[box=blue!6, below=of s] (q) {state $q$: members,\\pairs, team (20 values)};
\node[box=blue!14, below=of q] (m) {\textbf{\model}\\(\bestParams{} parameters)};
\node[box=orange!12, below=of m] (o) {forecasts, warnings,\\cue timing};
\draw[->] (s) -- (q); \draw[->] (q) -- (m); \draw[->] (m) -- (o);
\draw[->, orange!70!black] (o.east) -- ++(0.18,0) |- node[pos=0.25, right, align=left] {cue} (s.east);
\end{scope}
\end{tikzpicture}
\caption{\model{} in one picture. Left: the team (the ball) is pulled toward a resting point. Its height is the energy, regulation sets how fast it returns, and task progress moves the bowl. Right: \model{} turns headset data into forecasts, warnings and cue timing.}
\label{fig:teaser}
\end{figure}
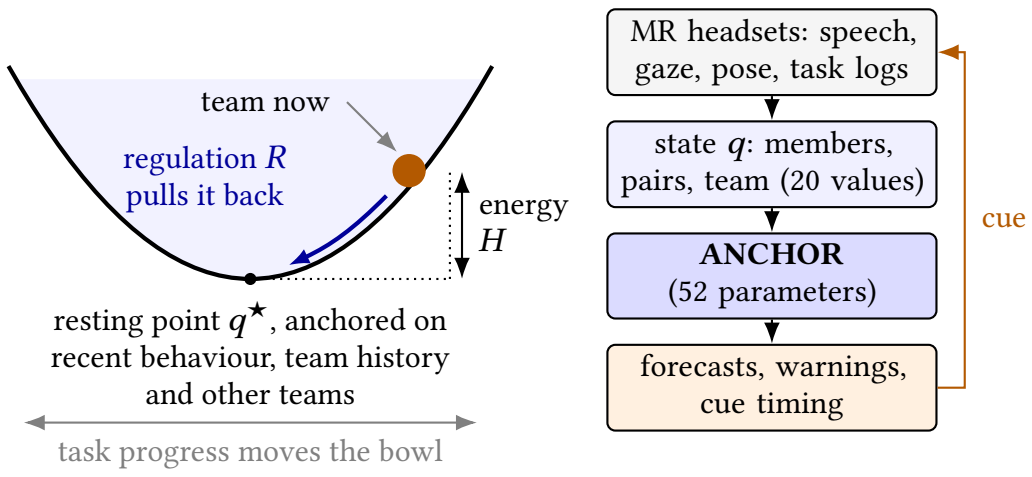

Collaborative mixed reality (MR) makes this loop practical. Each headset already senses who speaks, where its wearer looks and who stands together~\cite{murmr}, and it can deliver a light prompt, so the headsets act as sensors and the prompts as actuators (\cref{fig:teaser}). We study this loop with four-person teams that sort images together in MR. What the loop needs is a plant model. A controller needs a forecast several steps ahead that meets three demands. The model must learn from tens of teams, because each team is expensive to record. It must stay well behaved when iterated and when prompts are added. Finally, a designer must be able to read and question it.

\vspace{-1.5mm}
\paragraph{The idea.} Team psychology offers a simple picture. Small groups restore their balance after a disturbance~\cite{bales1953}, recurring social patterns act as attractors~\cite{vallacher1997,arrow2000}, and groups pull deviating members back to their norms~\cite{feldman1984}. This is how a ball behaves in a bowl: it rolls toward the bottom, it loses energy on the way, and the bowl can be moved from outside. We take this picture literally and give the team an energy, which is high when the team is far from its usual way of working and which regulation removes over time.

\emph{Port-Hamiltonian} (\pH) systems give this picture a precise form~\cite{vanderschaft2014,ortega2001}. A \pH{} model describes a system by the energy it stores, how that energy is dissipated, and the ports through which energy enters from outside, and it was first developed for physical systems such as circuits and robots. For teams, the energy is a modelling hypothesis, and we test it as one. We call the model \model{}, after its central element: a resting point \emph{anchored} on the team's recent behavior, its own history and the norms of other teams. The model replaces the question ``what comes next?'' with three questions that a person can reason about:

\begin{enumerate}
\item \textbf{Where does this team rest?}
\item \textbf{How does it return there?}
\item \textbf{What moves it away?}
\end{enumerate}

\paragraph{Contributions.}
\begin{itemize}[leftmargin=1em]
\item \textbf{A plant model of a human team.} To our knowledge, \model{} is the first port-Hamiltonian model of human team interaction. Each of its \bestParams{} parameters has a plain meaning, and each part stands for a principle of team psychology (\cref{sec:model}).
\item \textbf{A resting point with three anchors.} The model learns how much each behaviour follows the moment, the team and other teams (\cref{sec:rest,sec:teams}).
\item \textbf{Guarantees that hold by construction.} Every forecast obeys an energy balance across phase switches, and it stays bounded under any spaced schedule of bounded cues, for every parameter value (\cref{sec:guarantees}).
\item \textbf{A direct test on real teams.} We measure how much of a real deviation survives over time, against the model and shuffled sessions, and we remove each part of the model to test its principle on held-out teams (\cref{sec:real,sec:rq3}).
\item \textbf{A route to the closed loop.} With far fewer parameters than a black box of equal accuracy, the model supports early warning and cue timing, which we evaluate in simulation under assumed cue effects (\cref{sec:accuracy,sec:use}).
\end{itemize}

\section{The System and Its Data}\label{sec:state}

\paragraph{Sensing.} We use the MURMR dataset~\cite{murmr}, collected and described in recent  work in the literature~\cite{murmr,s2s}. \nGroups{} co-located teams of four, each member wearing a Meta Quest Pro, sorted 28 images into six emotion categories by discussion and consensus, with no time limit and no scripted turns. This image sorting task has been adopted in various group dynamics studies~\cite{s2s,murmr}. 
Every 16\,s we read a 32\,s window that records how long each member spoke, how long each pair's gaze hovered simultaneously on the same image, and how many seconds each pair stood within 0.46\,m on the floor. Sessions have \minWindows--\maxWindows{} windows (\nWindows{} in total). The attention and proximity signals are sparse, covering about 2\% and 4\% of a window on average.
\vspace{-2mm}
\paragraph{The state.} The state of a team at window $k$ is a vector $q_k$ of 20 numbers between 0 and 1. Four hold the speaking time of each member, six the shared attention of each pair and six the proximity of each pair. The last four are team-level quantities measured directly from the window: mean speaking, attention density, proximity density and \emph{speaking balance}. Balance is the normalized entropy of the speaking shares, so it is 1 when all four members speak equally and 0 when at most one speaks. \Cref{tab:state} lists the blocks. We keep team quantities as separate coordinates because a team state is more than its parts, and \cref{sec:rq3} shows that this matters.

% \begin{table}[t]
% \caption{The 20 coordinates of the state $q_k$ for one window. Every entry lies between 0 and 1.}
% \label{tab:state}
% \small
% \setlength{\tabcolsep}{3pt}
% \begin{tabular}{>{\raggedright\arraybackslash}p{1.7cm}c>{\raggedright\arraybackslash}p{5.5cm}}
% \toprule
% Block & Size & Each entry\\
% \midrule
% Speaking & 4 & Member's speaking time, as a share of the window\\
% Attention & 6 & Pair's simultaneous gaze on one image, as a share of the window\\
% Proximity & 6 & Pair's seconds within 0.46\,m, as a share of the window\\
% Team & 4 & Mean speaking, attention density, proximity density, speaking balance\\
% \bottomrule
% \end{tabular}
% \end{table}
\begin{table}[t]
\caption{The 20 numbers of the state $q_k$ for one window. Each entry is a share of the window, so it lies between 0 and 1.}
\label{tab:state}
\small
\setlength{\tabcolsep}{4pt}
\begin{tabular}{@{}lcl@{}}
\toprule
Block & Size & Each entry\\
\midrule
Speaking & 4 & Member's speaking time\\
Attention & 6 & Pair's simultaneous gaze on one image\\
Proximity & 6 & Pair's time within 0.46\,m\\
Team & 4 & Mean of each behavior; speaking balance\\
\bottomrule
\end{tabular}
\end{table}

\vspace{-2mm}
\paragraph{Task progress.} The MR application keeps \emph{task logs} of who places which image in which category, and when. From them we compute the \emph{remaining work} $\pi_k\in[0,1]$, the share of the 28 images not yet decided, where an image counts as decided once its label has stayed unchanged for 64,s. It uses only information available at window $k$, so a live system can compute it.
\vspace{-2mm}
\paragraph{The goal.} From training teams we want to learn a map that takes the history of a new team up to window $k$ and forecasts its state $\hat q_{k+h}$ for $h=1,\dots,16$ (16--256\,s ahead). The forecast must stay bounded when iterated and when cues are applied.

\section{The \model{} Model}\label{sec:model}

The model rests on six principles from team psychology.
\begin{itemize}[leftmargin=*, topsep=1pt]
\item[\textbf{P1}] \textbf{Teams have a resting point.} Interaction settles into recurrent patterns, described as group equilibria~\cite{bales1953}, attractors~\cite{vallacher1997,arrow2000} or set points~\cite{gottman2002}.
\item[\textbf{P2}] \textbf{Deviations are regulated back.} After a disturbance, such as one member dominating, groups restore their pattern~\cite{bales1953,feldman1984}.
\item[\textbf{P3}] \textbf{Team states emerge across levels.} Team-level properties arise from members and pairs, but they are not functions of them~\cite{kozlowski2000,kozlowski2013}.
\item[\textbf{P4}] \textbf{Teams move through phases.} Exploration and consolidation alternate or follow each other~\cite{gersick1988,marks2001}.
\item[\textbf{P5}] \textbf{The task drives interaction.} Coordination changes as decisions settle~\cite{marks2001}.
\item[\textbf{P6}] \textbf{Interventions must be timed.} Coaching and cues work when delivered at the right moment~\cite{hackman2005}.
\end{itemize}

\Cref{tab:mapping} maps each principle onto one part of the model, which makes each one a hypothesis that held-out teams can confirm or reject. Readers from circuits or robotics may find the analogy of \cref{tab:analogy} in \cref{app:model} helpful. The next subsections answer the three questions in turn: where the team rests (\cref{sec:rest}), how it returns (\cref{sec:return}) and what moves it (\cref{sec:moves}), and \cref{tab:notation} in \cref{app:notation} lists every symbol.

\begin{table*}[t]
\caption{From team behaviour to \model{}, in plain terms. The principles P1--P6 and their sources are listed above, and \cref{tab:principles} gives the test of each part on held-out teams.}
\label{tab:mapping}
\small
\setlength{\tabcolsep}{4pt}
\begin{tabular*}{\textwidth}{@{\extracolsep{\fill}}llll@{}}
\toprule
Question & Part of the model & What it means for a team & Sensor source\\
\midrule
\textbf{Where does it rest?} & Resting point $q^\star$ (P1) & The pattern it returns to, anchored on its past and other teams & Sociogram history\\
& State across levels $q$ (P3) & Team states emerge from members and pairs but are not reducible to them & Audio, gaze, head pose\\
\midrule
\textbf{How does it return?} & Energy $H$, regulation $R_\sigma$ (P2) & Distance from the pattern, and how fast deviations are regulated back & ---\\
& Link gains in $K$, $R_\sigma$, $J_\sigma$ & Who interacts with whom constrains coordination & Team structure\\
\midrule
\textbf{What moves it away?} & Task input $g\,u$ (P5) & Task progress shifts the pattern & Task logs\\
& Phase $\sigma$ (P4) & Exploration and consolidation regulate differently & Task logs\\
& Cue $d$ (P6) & A well-timed prompt moves the team & Delivered by the system\\
\bottomrule
\end{tabular*}
\end{table*}

%\subsection{Where does the team rest?}\label{sec:rest}

\subsection{Resting Point and Energy}\label{sec:rest}
The \emph{resting point} $q^\star$ is the interaction pattern a team would settle into if nothing new happened. At any moment the team is somewhere else, because one member may be dominating or two members may have just huddled around an image. We write $z=q-q^\star$ for this deviation and measure it with an energy:
\begin{equation}
H=\tfrac12\,z^\top K\,z.\label{eq:H}
\end{equation}
The matrix $K$ is the \emph{stiffness}. It is positive definite, so $H$ is a true bowl whose only lowest point is $q^\star$, and a larger $K$ makes the bowl steeper.

Each team has its own way of working, so \model{} computes the resting point at every forecast start as a blend of three anchors:

\begin{equation}
q^\star=\ann{w_{\mathrm{rec}}\odot\bar q_{\mathrm{rec}}}{the moment}+\ann{w_{\mathrm{ses}}\odot\bar q_{\mathrm{ses}}}{this team}+\ann{w_{\mathrm{pop}}\odot\bar q_{\mathrm{pop}}}{other teams}.\label{eq:attractor}
\end{equation}
Here $\bar q_{\mathrm{rec}}$ is a weighted average of the last five windows (about 80\,s), $\bar q_{\mathrm{ses}}$ is the average of the session so far, and $\bar q_{\mathrm{pop}}$ is the average of the training teams or population. The symbol $\odot$ multiplies element by element, and for each kind of behavior the three learned weights are non-negative and sum to one. Each anchor is a hypothesis: teams persist in what they just did ($\bar q_{\mathrm{rec}}$), each team has its own style ($\bar q_{\mathrm{ses}}$), and teams doing the same task share norms ($\bar q_{\mathrm{pop}}$). The learned weights show which hypothesis holds for which behavior.

\begin{figure}[t]
\centering
\includegraphics[width=\columnwidth]{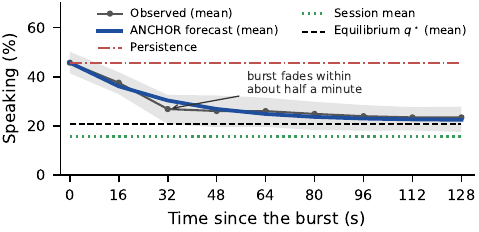}
\caption{Speaking bursts fade, and \model{} predicts how. Curves average all \burstN{} bursts on held-out teams in which a member speaks at least \burstGap{} points above their session average. % (band: 95\% interval over teams). 
\model{} (blue), fitted without the team, follows the decay, whereas persistence assumes that the burst continues and the session mean assumes that it never happened.}
\label{fig:attractor}
\end{figure}

%\paragraph{An example on real data.} \Cref{fig:attractor} averages all bursts of speech on held-out teams. At the burst the member speaks \burstNow\% of the time on average, and 32\,s later this has fallen to \burstObsTwo\%, against a forecast of \burstPredTwo\% from \model{}, which was fitted without that team. The burst is a ball pushed up the bowl, and the data show it rolling back.

% \paragraph{A first look at real data.} Before the formal evaluation, we check the bowl picture on a simple event, a member who suddenly speaks much more than usual. \Cref{fig:attractor} averages all \burstN{} such bursts on held-out teams, with \model{} fitted without each team. At the burst the member speaks \burstNow\% of the time on average, and 32\,s later this has fallen to \burstObsTwo\%, against a forecast of \burstPredTwo\%. The fitted resting point averages \burstQstar\%, above the session mean of \burstSession\%, because \model{} gives the recent windows a weight of 0.44. The fitted recovery time of 26--37\,s sets how fast the burst fades. Over the next two minutes the forecast misses by \burstMaePh{} points on average, against \burstMaeSes{} for the session mean and \burstMaePers{} for persistence. The burst is a ball pushed up the bowl, and the data show it rolling back at about the predicted rate. \Cref{sec:results} tests this across all behaviors and horizons.

\vspace{-2mm}
\paragraph{A first look at real data.} Before the formal evaluation, we test the bowl picture on a simple event, a member who suddenly speaks much more than usual. \Cref{fig:attractor} averages all \burstN{} such bursts on held-out teams, with \model{} fitted without each team. Speaking falls from \burstNow\% at the burst to \burstObsTwo\% 32\,s later, against a forecast of \burstPredTwo\%, so the ball pushed up the bowl rolls back at about the predicted rate. Over two minutes the forecast error is \burstMaePh{} points, against \burstMaeSes{} for the session mean and \burstMaePers{} for persistence. \Cref{sec:results} tests this across all behaviors.

%\subsection{How does the team return?}\label{sec:return}
\subsection{Regulation and Exchange}\label{sec:return}
The energy says how far the team is from rest, and the dynamics say how it moves. A \pH{} system describes them by splitting the motion into parts with clear roles~\cite{vanderschaft2014}:

\begin{equation}
\dot q=\big(\ann{J_\sigma}{exchange}-\ann{R_\sigma}{regulation}\big)Kz+\ann{g\,u}{task input}.\label{eq:ph}
\end{equation}
The vector $Kz$ is the gradient of the energy, which points uphill in the bowl. The subscript $\sigma$ is the team's current phase, exploration or consolidation (P4), and $g\,u$ is the push that task progress gives; \cref{sec:moves} says how both are computed.

\emph{Regulation} $R_\sigma$ is positive definite and removes energy. It is the term that brings a team back, and a deviation in one coordinate fades with a time constant of about $16\,\mathrm{s}/(R_{ii}K_{ii})$\footnote{The model counts time in windows, and a new window starts every 16\,s. For one coordinate on its own, \cref{eq:ph} reduces to $\dot z_i=-R_{ii}K_{ii}\,z_i$, whose time constant is $1/(R_{ii}K_{ii})$ windows, that is, $16\,\mathrm{s}/(R_{ii}K_{ii})$. The value is approximate because coordinates are coupled through $K$ and $R_\sigma$ and the forecast advances in whole steps.}, so its entries read directly as recovery times. \emph{Exchange} $J_\sigma$ is skew-symmetric ($J_\sigma=-J_\sigma^\top$), so it moves energy between coordinates without changing the total. It could describe talk turning into shared focus, and we keep it as a hypothesis to test.

The energy shows why these roles are clean. Differentiating \cref{eq:H} along \cref{eq:ph} with $q^\star$ fixed gives
\begin{equation}
\dot H=(Kz)^\top\dot q=\ann{-(Kz)^\top R_\sigma(Kz)}{regulation removes}+\ann{(Kz)^\top g\,u}{the task supplies},\label{eq:Hdot}
\end{equation}
since the exchange term $(Kz)^\top J_\sigma(Kz)$ vanishes for a skew-symmetric $J_\sigma$. Without task input the ball can only go down. This one line is the seed of the guarantees in \cref{sec:guarantees}, and \cref{tab:mapping} lists the principle that each part stands for.

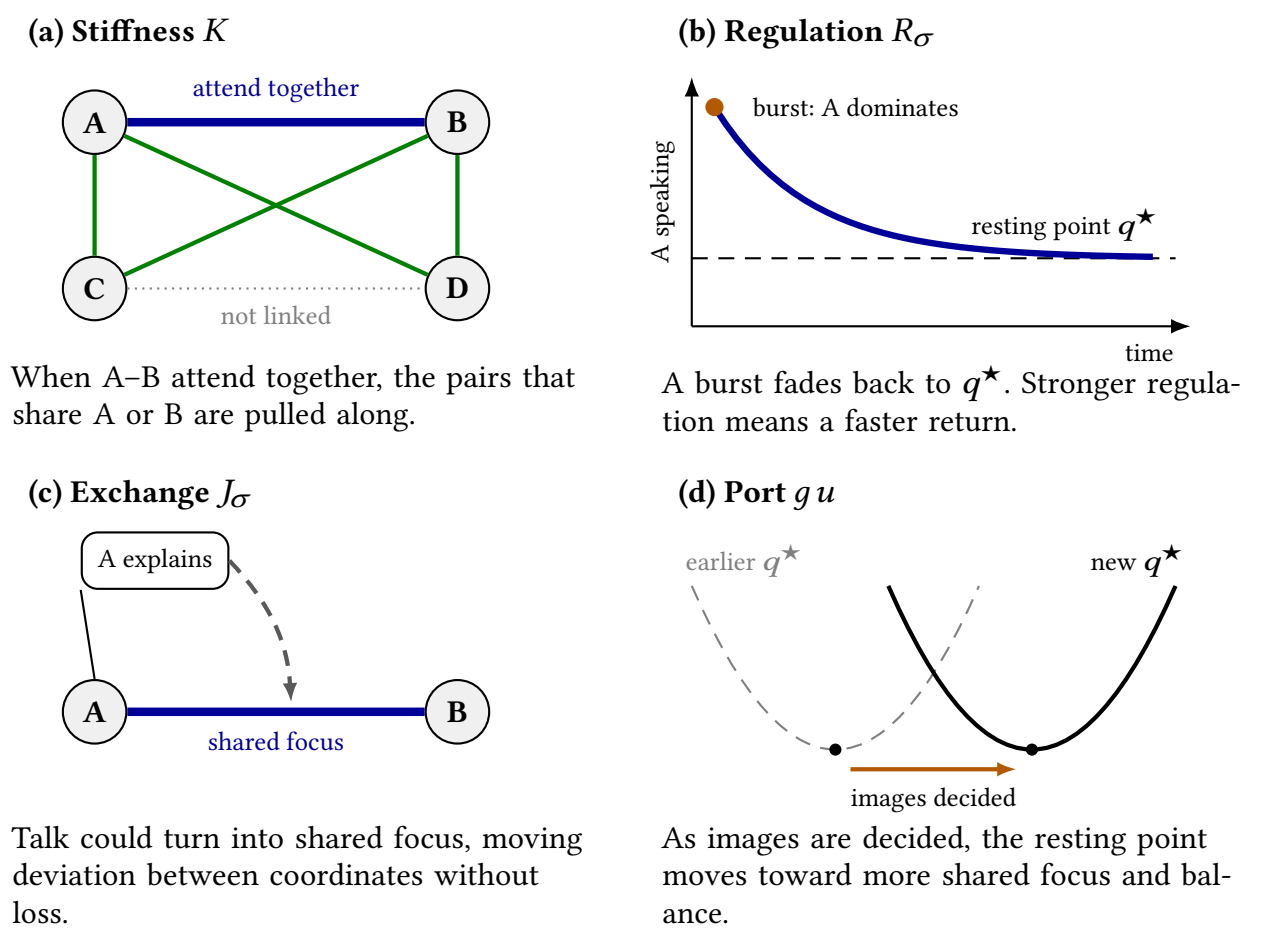
\begin{figure}[t]
\centering
\begin{tikzpicture}[font=\scriptsize, >={Latex[length=1.3mm]},
  person/.style={circle, draw, fill=gray!12, inner sep=0pt, minimum size=0.42cm, font=\scriptsize\bfseries},
  ttl/.style={anchor=north west, font=\scriptsize\bfseries},
  vcap/.style={anchor=north west, text width=3.85cm, align=left, font=\scriptsize, inner sep=0pt}]
\begin{scope}
\node[ttl] at (0,2.25) {(a) Stiffness $K$};
\node[person] (A) at (0.55,1.45) {A}; \node[person] (B) at (2.95,1.45) {B};
\node[person] (C) at (0.55,0.35) {C}; \node[person] (D) at (2.95,0.35) {D};
\draw[blue!60!black, line width=1.6pt] (A) -- node[above=0pt, font=\tiny] {attend together} (B);
\foreach \u/\v in {A/C,A/D,B/C,B/D} \draw[green!50!black, line width=0.8pt] (\u) -- (\v);
\draw[gray, densely dotted] (C) -- node[below=0pt, font=\tiny] {not linked} (D);
\node[vcap] at (0,-0.15) {When A--B attend together, the pairs that share A or B are pulled along.};
\end{scope}
\begin{scope}[xshift=4.3cm]
\node[ttl] at (0,2.25) {(b) Regulation $R_\sigma$};
\draw[->] (0.2,0.1) -- (3.5,0.1) node[below left=0pt, font=\tiny] {time};
\draw[->] (0.2,0.1) -- (0.2,1.75) node[midway, left=0pt, rotate=90, anchor=south, font=\tiny] {A speaking};
\draw[densely dashed] (0.2,0.55) -- (3.4,0.55) node[above left=0pt, font=\tiny] {resting point $q^\star$};
\draw[blue!60!black, line width=1.2pt, domain=0:2.9, samples=40] plot ({0.35+\x}, {0.55+1.0*exp(-1.6*\x)});
\fill[orange!70!black] (0.35,1.55) circle (0.06) node[right=4pt, font=\tiny, black] {burst: A dominates};
\node[vcap] at (0,-0.15) {A burst fades back to $q^\star$. Stronger regulation means a faster return.};
\end{scope}
\begin{scope}[yshift=-3.05cm]
\node[ttl] at (0,2.25) {(c) Exchange $J_\sigma$};
\node[person] (A2) at (0.55,0.6) {A}; \node[person] (B2) at (2.95,0.6) {B};
\draw[blue!60!black, line width=1.6pt] (A2) -- node[below=0pt, font=\tiny] {shared focus} (B2);
\node[draw, rounded corners=3pt, fill=white, font=\tiny] (talk) at (0.95,1.6) {A explains};
\draw (A2.north) -- (talk.south west);
\draw[->, densely dashed, thick, gray!70!black] (talk.east) to[bend left=20] (1.85,0.66);
\node[vcap] at (0,-0.15) {Talk could turn into shared focus, moving deviation between coordinates without loss.};
\end{scope}
\begin{scope}[xshift=4.3cm, yshift=-3.05cm]
\node[ttl] at (0,2.25) {(d) Port $g\,u$};
\draw[gray, densely dashed, domain=-0.95:0.95, samples=30] plot ({1.15+\x}, {0.35+1.2*\x*\x});
\draw[thick, domain=-0.95:0.95, samples=30] plot ({2.45+\x}, {0.35+1.2*\x*\x});
\fill (1.15,0.35) circle (0.04); \fill (2.45,0.35) circle (0.04);
\draw[->, thick, orange!70!black] (1.25,0.22) -- node[below=0pt, font=\tiny, black] {images decided} (2.35,0.22);
\node[font=\tiny, gray] at (0.55,1.6) {earlier $q^\star$};
\node[font=\tiny] at (3.15,1.6) {new $q^\star$};
\node[vcap] at (0,-0.15) {As images are decided, the resting point moves toward more shared focus and balance.};
\end{scope}
\end{tikzpicture}
\caption{What each term means for a team of four (A--D). Each kind of link in (a) has a single learned gain, which keeps the model small and independent of how members are labelled.}
\label{fig:matrices}
\end{figure}
\vspace{-2mm}
\paragraph{From team structure to matrices.} The state has 20 coordinates, so $K$, $R_\sigma$ and $J_\sigma$ are each $20\times20$, and learning every entry would take hundreds of parameters per matrix. It would also make the forecast depend on who happens to be called member ``A''. We instead build them from the team's structure, so that all coordinates playing the same role share one number (\cref{fig:matrices}a). These relations stay the same when members are renamed, so the model treats the four peers alike. We model the matrices as follows.

%%%% Stiffness K ------ 

% Inside the stiffness, the idea is that related coordinates tend to move together. If A and B attend together, the other pairs that include A or B are pulled along (\cref{fig:matrices}a), and a member's speaking is tied in the same way to the pairs that include them. We encode five such links: (1) the attention and proximity of the same pair, (2) attention pairs that share a member, (3) proximity pairs that share a member, (4) a speaker and the attention pairs containing them, and (5) a speaker and the proximity pairs containing them. Each link makes it cost energy when two linked coordinates deviate differently, adding $\tfrac12\gamma_\ell(z_a-z_b)^2$ for each linked pair of coordinates $a$ and $b$, with one learned gain $\gamma_\ell\ge0$ per kind of link. Summed over one kind of link, these terms form the graph Laplacian $L_\ell$ of that link, so for the 16 member and pair coordinates the block $K_d$ of $K$ is
% \begin{equation}
% K_d=\mathrm{diag}\big(e^{\kappa_S}\mathbf 1_4,\ e^{\kappa_A}\mathbf 1_6,\ e^{\kappa_P}\mathbf 1_6\big)+\sum_{\ell=1}^{5}\gamma_\ell L_\ell,\label{eq:Kd}
% \end{equation}
% where the exponentials keep each behaviour's own steepness positive. A last gain links each team coordinate to the mean of its behaviour, which ties the team level to the levels below it. Each term is positive semidefinite and the diagonal is positive, so $K$ is positive definite for every value of these gains, which is what makes $H$ a true bowl without any constraint on training.
\vspace{-2mm}
\paragraph{The stiffness matrix $K$} Inside the stiffness, the idea is that related coordinates tend to move together. If A and B attend together, the other pairs that include A or B are pulled along (\cref{fig:matrices}a), and a member's speaking is tied in the same way to the pairs that include them. We encode five such links: the attention and proximity of the same pair, attention pairs that share a member, proximity pairs that share a member, a speaker and the attention pairs containing them, and a speaker and the proximity pairs containing them. Each link makes it cost energy when two linked coordinates deviate differently, adding $\tfrac12\gamma_\ell(z_a-z_b)^2$ for each linked pair of coordinates $a$ and $b$, with one learned gain $\gamma_\ell\ge0$ per kind of link. Summed over one kind of link, these terms form the graph Laplacian $L_\ell$ of that link,\footnote{Take two linked coordinates $a$ and $b$ with the same steepness $k$. Without the link, the energy is $\tfrac12 k z_a^2+\tfrac12 k z_b^2$, so each coordinate is pulled back on its own. The link adds $\tfrac12\gamma(z_a-z_b)^2$, which is zero when $a$ and $b$ deviate by the same amount and grows when they deviate differently, so it pulls them toward each other. Written as $\tfrac12 z^\top Kz$, this energy has $K=\left(\begin{smallmatrix}k+\gamma&-\gamma\\-\gamma&k+\gamma\end{smallmatrix}\right)$, and the off-diagonal $-\gamma$ is the link.} and the full stiffness matrix $K$ is
\begin{equation}
K=\begin{pmatrix}\mathrm{diag}\big(e^{\kappa_S}\mathbf 1_4,\,e^{\kappa_A}\mathbf 1_6,\,e^{\kappa_P}\mathbf 1_6\big)+\textstyle\sum_{\ell=1}^{5}\gamma_\ell L_\ell & 0\\ 0 & \mathrm{diag}\big(e^{\kappa_T}\big)\end{pmatrix}\!+\chi L_{\mathrm{cons}}.\label{eq:Kd}
\end{equation}

The top-left block acts on the 16 member and pair coordinates. Its first term gives each coordinate its own steepness, with one value shared by the four speaking coordinates ($e^{\kappa_S}$), one by the six attention pairs ($e^{\kappa_A}$) and one by the six proximity pairs ($e^{\kappa_P}$); $\mathbf 1_n$ is a vector of $n$ ones, and the exponentials keep these values positive. Its second term adds the five links. The bottom-right block gives each of the four team coordinates its own steepness, the four entries of $e^{\kappa_T}$. The last term ties the team coordinates to the member and pair coordinates with one gain $\chi\ge0$. For speaking, attention and proximity it adds $\tfrac12\chi(t-m)^2$, where $t$ is the team coordinate of that behavior and $m$ is the mean of its member or pair coordinates, so the team level stays close to what its members and pairs show. Speaking balance is an entropy rather than an average, so it is left out of this tie. Counting $\kappa_S,\kappa_A,\kappa_P$, the five $\gamma_\ell$, the four team steepnesses and $\chi$ gives the 13 stiffness values of \cref{tab:params}. Each term is positive semidefinite and the diagonal is positive, so $K$ is positive definite for every value of these gains, which is what makes $H$ a true bowl without any constraint on training.

%%%% Regulation R ------ 

% Regulation is simpler. $R_\sigma$ is diagonal with one positive value per kind of behaviour and per team coordinate in each phase, so comparing its two phases directly asks whether a team recovers faster when consolidating than when exploring. Exchange has one gain for speaking and attention and one for attention and proximity, again in each phase, and its skew-symmetric form is what prevents it from adding or removing energy.
\vspace{-2mm}
\paragraph{The regulation matrix $R_\sigma$} Regulation is simpler, since it has no links:
\begin{equation}
R_\sigma=\mathrm{diag}\big(e^{\zeta_{\sigma,S}}\mathbf 1_4,\,e^{\zeta_{\sigma,A}}\mathbf 1_6,\,e^{\zeta_{\sigma,P}}\mathbf 1_6,\,e^{\zeta_{\sigma,T}}\big).\label{eq:R}
\end{equation}
Each coordinate gets its own recovery rate, with one value shared by the four speaking coordinates ($e^{\zeta_{\sigma,S}}$), one by the six attention pairs ($e^{\zeta_{\sigma,A}}$) and one by the six proximity pairs ($e^{\zeta_{\sigma,P}}$), while each of the four team coordinates has its own value, the four entries of $e^{\zeta_{\sigma,T}}$. The exponentials keep these values positive, so $R_\sigma$ is positive definite. Each phase $\sigma$ has its own set of seven values, which gives the 14 regulation values of \cref{tab:params}, and comparing the two sets directly asks whether a team recovers faster when consolidating than when exploring. 

%%%% Exchange J ------ 

% \paragraph{The exchange matrix $J_\sigma$} Exchange has one gain for speaking and attention and one for attention and proximity, again in each phase, and its skew-symmetric form is what prevents it from adding or removing energy.

% \paragraph{The exchange matrix $J_\sigma$} Exchange links different behaviors rather than coordinates of the same kind:
% \begin{equation}
% J_\sigma=\begin{pmatrix}j_{\sigma,1}S_{SA}+j_{\sigma,2}S_{AP} & 0\\ 0 & 0\end{pmatrix}.\label{eq:J}
% \end{equation}
% Here $S_{SA}$ has $+1$ in the row of each speaker and the column of each attention pair containing them, and $-1$ in the mirrored entry, so it can move energy between a member's speaking and the shared attention of their pairs. $S_{AP}$ does the same between the attention and the proximity of each pair. The gains $j_{\sigma,1}$ and $j_{\sigma,2}$ can take either sign, which sets the direction of the flow. The team block is zero, so exchange acts only on members and pairs. Because $S_{SA}$ and $S_{AP}$ are skew-symmetric, so is $J_\sigma$, which is what prevents it from adding or removing energy. Each phase has its own two gains, which gives the 4 exchange values of \cref{tab:params}.

\paragraph{The exchange matrix $J_\sigma$} Exchange links different behaviors rather than coordinates of the same kind:
\begin{equation}
J_\sigma=\begin{pmatrix}j_{\sigma,SA}S_{SA}+j_{\sigma,AP}S_{AP} & 0\\ 0 & 0\end{pmatrix}.\label{eq:J}
\end{equation}
Here $S_{SA}$ has $+1$ in the row of each speaker and the column of each attention pair containing them, and $-1$ in the mirrored entry, so it can move energy between a member's speaking and the shared attention of their pairs. $S_{AP}$ does the same between the attention and the proximity of each pair. The gains $j_{\sigma,SA}$ and $j_{\sigma,AP}$ can take either sign, which sets the direction of the flow. The team block is zero, so exchange acts only on members and pairs. Because $S_{SA}$ and $S_{AP}$ are skew-symmetric, so is $J_\sigma$, which is what prevents it from adding or removing energy. Each phase has its own two gains, which gives the 4 exchange values of \cref{tab:params}. These two links encode the \emph{hypotheses} that talk draws members into shared focus and that standing together enables it. %Exchange between interchangeable members would break their symmetry, and the richer families of \cref{tab:ablation}, whose exchange can link any two behaviours, did not clearly improve forecasts.

% \Cref{tab:params} shows that this leaves \bestParams{} parameters, each with a name and a meaning, whereas a free symmetric $K$ alone would need $20\cdot21/2=210$ values. The full construction is in \cref{app:model}.

% \begin{table}[t]
% \caption{Where the \bestParams{} free parameters come from, with the symbols of \cref{eq:Kd,eq:R,eq:J}. Each value is shared by all coordinates of one behaviour (speaking, attention or proximity) or belongs to one of the four team coordinates; regulation and exchange have a separate set in each of the two phases.}
% \label{tab:params}
% \small
% \setlength{\tabcolsep}{3pt}
% \begin{tabular}{>{\raggedright\arraybackslash}p{1.55cm}>{\raggedright\arraybackslash}p{2.35cm}>{\raggedright\arraybackslash}p{3.25cm}r}
% \toprule
% Part & Symbols & One value per\ldots & Count\\
% \midrule
% Stiffness $K$ & $\kappa_S,\kappa_A,\kappa_P$ & behaviour & 3\\
% & $\gamma_1,\dots,\gamma_5$ & kind of link & 5\\
% & $\kappa_T$ & team coordinate & 4\\
% & $\chi$ & tie between team and member or pair coordinates & 1\\
% Regulation $R_\sigma$ & $\zeta_{\sigma,S},\zeta_{\sigma,A},\zeta_{\sigma,P}$, $\zeta_{\sigma,T}$ & behaviour (3) and team coordinate (4), in each phase (2) & 14\\
% Exchange $J_\sigma$ & $j_{\sigma,SA},j_{\sigma,AP}$ & link (2), in each phase (2) & 4\\
% Port $g=Kb$ & $b$ & behaviour (3) and team coordinate (4) & 7\\
% Anchor weights & $w_{\mathrm{rec}},w_{\mathrm{ses}},w_{\mathrm{pop}}$ & behaviour or team coordinate (7), 2 free of 3 because they sum to one & 14\\
% \midrule
% \textbf{Total} & & & \textbf{\bestParams}\\
% \bottomrule
% \end{tabular}
% \end{table}

\begin{table}[t]
\caption{Where the \bestParams{} free parameters come from, with the symbols of \cref{eq:Kd,eq:R,eq:J}. Each value is shared by all coordinates of one behavior (3) or belongs to one team coordinate (4), regulation and exchange have one set per phase (2), and 2 of the 3 anchor weights are free because they sum to one.}
\label{tab:params}
\small
\setlength{\tabcolsep}{4pt}
\begin{tabular}{@{}llr@{}}
\toprule
Part & Symbols (values) & Count\\
\midrule
Stiffness $K$ & $\kappa_S,\kappa_A,\kappa_P$ (3), $\gamma_1..\gamma_5$ (5), $\kappa_T$ (4), $\chi$ (1) & 13\\
Regulation $R_\sigma$ & $\zeta_{\sigma,S},\zeta_{\sigma,A},\zeta_{\sigma,P},\zeta_{\sigma,T}$ & $(3{+}4)\times2=14$\\
Exchange $J_\sigma$ & $j_{\sigma,SA},j_{\sigma,AP}$ & $2\times2=4$\\
Port $g=Kb$ & $b$ & $3+4=7$\\
Anchors & $w_{\mathrm{rec}},w_{\mathrm{ses}},w_{\mathrm{pop}}$ & $(3{+}4)\times2=14$\\
\midrule
\textbf{Total} & & \textbf{\bestParams}\\
\bottomrule
\end{tabular}
\end{table}

%\subsection{What moves the team?}\label{sec:moves}
\subsection{Task, Phase, and Cues}\label{sec:moves}

% Three inputs can move the team. \emph{Task progress} enters through the port $g\,u$ in \cref{eq:ph}, where $u_h$ is the forecast change in remaining work $h$ steps ahead and $g$ is learned. A steady input does not push the team forever, since it moves the bottom of the bowl and the team then follows at its own pace (\cref{fig:matrices}d). \emph{Phases} change how the team regulates. We use two values of $\sigma$: exploration while at least half of the images are expected to be undecided, and consolidation otherwise. All phases share the same energy $H$, because switching between modes that are each stable can be unstable unless they share one energy~\cite{liberzon2003}. A \emph{cue}, such as a prompt that invites a quiet member to speak, is a kick that moves the deviation, $z\leftarrow z+d$, and the model then predicts how long its effect lasts.

Three inputs can move the team. \emph{Task progress} enters through the port $g\,u$ in \cref{eq:ph}, where $u_h$ is the forecast change in remaining work $h$ steps ahead and $g=Kb$ sets how strongly task progress pushes each coordinate, with one learned value of $b$ per behavior and one per team coordinate (7 in all). A steady input moves the bottom of the bowl, and the team then follows at its own pace (\cref{fig:matrices}d). \emph{Phases} change how the team regulates. We use two values of $\sigma$: exploration while at least half of the images are expected to be undecided, and consolidation otherwise. All phases share the same energy $H$, because a shared energy keeps switching between stable modes stable~\cite{liberzon2003}. A \emph{cue}, such as a prompt that invites a quiet member to speak, is a kick that moves the deviation, $z\leftarrow z+d$, and the model then predicts how long its effect lasts.

With the port, every part of the model is now in place. \Cref{tab:params} counts its \bestParams{} parameters, each with a name and a meaning. A free symmetric $K$ alone would need $210$ values\footnote{A symmetric $n\times n$ matrix has $n(n+1)/2$ free entries.} while \model{} uses 13 parameters for $K$. The full construction is in \cref{app:model}.

%\subsection{From model to forecast}\label{sec:forecast}
\subsection{Forecasting and Learning} \label{sec:forecast}

A forecast predicts where the team will be in each of the next windows if nothing new happens apart from the task. \model{} computes it by rolling the ball forward one window at a time. It starts from the current deviation $z_0=q_k-q^\star$, with $q^\star$ and all matrices held fixed for that forecast. At each step, regulation shrinks the deviation, exchange moves part of it between behaviors and the task pushes it, and the forecast is the resting point plus whatever deviation is left.

Each step uses the implicit midpoint rule, which evaluates the motion halfway through the step rather than at its start. This keeps the energy balance of \cref{eq:Hdot} exact, whereas a plain Euler step can add energy that the model does not have~\cite{hairer2006}. Because $q^\star$ is fixed, \cref{eq:ph} reads $\dot z=A_\sigma z+g\,u$ with $A_\sigma=(J_\sigma-R_\sigma)K$, which is linear in $z$, so each step is a matrix multiplication:\footnote{Applying the midpoint rule over one window gives $z_{h+1}-z_h=A_\sigma\tfrac12(z_h+z_{h+1})+g\bar u_h$, and solving for $z_{h+1}$ gives \cref{eq:step}.}

\begin{equation}
z_{h+1}=\ann{C_\sigma z_h}{deviation left}+\ann{G_\sigma\bar u_h}{task push},\quad \hat q_{k+h}=\ann{q^\star}{resting point}+\ann{z_h}{deviation left}.\label{eq:step}
\end{equation}
Without task input, $z_{h+1}=C_\sigma z_h$, so $C_\sigma$ gives the part of the current deviation that is left one window later. The second term adds the task's push, since $G_\sigma$ turns the task input $\bar u_h$ during the window into a change of each coordinate. Explicitly, $C_\sigma=(I-\tfrac12A_\sigma)^{-1}(I+\tfrac12A_\sigma)$ and $G_\sigma=(I-\tfrac12A_\sigma)^{-1}g$, and $\bar u_h=\tfrac12(u_h+u_{h+1})$ is the task input halfway through the step. For one coordinate on its own, $C_\sigma$ reduces to the factor $(1-\tfrac a2)/(1+\tfrac a2)$ with $a=R_{ii}K_{ii}$, which is close to $e^{-a}$ when $a$ is small. Far ahead, every forecast settles at the bottom of the bowl, so the resting point sets long-range accuracy and the dynamics set the path there.

\vspace{-2mm}
\paragraph{Learning.} The parameters are chosen so that forecasts match what the training teams actually did over the next few windows. We learn all parameters $\theta$ by minimizing the error of forecasts up to a training horizon of $h_{\mathrm{tr}}$ windows,
\begin{equation}
\mathcal L(\theta)=\sum_{k}\sum_{h=1}^{h_{\mathrm{tr}}}\big\|\hat q_{k+h}(\theta)-q_{k+h}\big\|_{\nu}^{2}\ ,\label{eq:loss}
\end{equation}
% where $\|x\|_\nu^2=\sum_c\nu_cx_c^2$. The weights matter because the behaviors live on very different scales. Shared attention needs two members' gaze on the same image at the same moment, so it fills only about 2\% of a window on average, against about 10\% for a member's speaking, and its variance is about 14 times smaller (calculated from MURMR dataset~\cite{murmr}). With a plain squared error, a forecast that ignored attention would cost little, and the fit would serve speaking and proximity. Setting $\nu_c$ to the inverse training variance of coordinate $c$'s behavior divides each error by that behavior's typical spread, so every behavior counts equally. %, as in the nMSE of \cref{sec:eval}.
% No energy or stability term appears in the loss, because the structure already guarantees these properties.
where $\|x\|_\nu^2=\sum_c\nu_cx_c^2$. The weights matter because the behaviors live on very different scales. Shared attention needs two members' gaze on the same image at the same moment, so it fills only about 2\% of a window on average, against about 10\% for a member's speaking, and its variance is about 14 times smaller (calculated from MURMR dataset~\cite{murmr}). With a plain squared error, a forecast that ignored attention would cost little, and the fit would serve speaking and proximity. Setting $\nu_c$ to the inverse training variance of coordinate $c$'s behavior divides each error by that behavior's typical spread, so every behavior counts equally.%, as in the nMSE of \cref{sec:eval}. 
The loss needs no energy or stability term, because the structure already guarantees these properties.

\section{Guarantees for Closed-Loop Use}\label{sec:guarantees}

%A forecaster that is iterated on its own output can diverge, and a planner built on it can then justify anything. In \model{} this cannot happen, and the reason lies in how we construct the matrices in \cref{sec:model}.% rather than in the fitted values. We state the two results here, and \cref{app:proofs} gives the formal proofs.
A forecaster that is iterated on its own output must stay bounded for a planner built on it to be trustworthy. \model{} stays bounded by construction, because of the form of its matrices rather than their fitted values. We state the two results here, with formal proofs in \cref{app:proofs}, and then show how the model runs in a live loop.

%%% In the tech report this is the discrete time passivity 
\begin{proposition}[Energy balance]\label{prop:passivity}
For every phase sequence and every input, each forecast step satisfies
\begin{equation}
H_{h+1}-H_h=\ann{-v^\top R_\sigma v}{removed, $\le0$}+\ann{v^\top g\,\bar u_h}{supplied by the task},\label{eq:balance}
\end{equation}
where $v=\tfrac12K(z_h+z_{h+1})$ is the energy gradient $Kz$ of \cref{eq:Hdot} at the midpoint of the step.
\end{proposition}

% In words, regulation can only remove energy, exchange cannot change it, and only the task can add it. Without task input, the forecast therefore stays inside the bowl where it started, however far it runs and however often the phase switches. The proof is the discrete form of \cref{eq:Hdot}: the midpoint rule makes the step use one gradient $v$, so the exchange term vanishes exactly, and only the current step's matrices appear, which is why any phase sequence is allowed. 

% Because every regulation value is positive, each step without task input also removes a positive share of the energy in every direction. We measure the distance from rest with the energy norm $\|z\|_K=\sqrt{z^\top Kz}=\sqrt{2H}$, built from the energy of \cref{eq:H}. In each step this norm shrinks by at least the factor $\rho=\max_\sigma\|C_\sigma\|_K<1$, the largest fraction of this distance that can remain after one step (\cref{lem:contraction} in \cref{app:proofs}).

% In words, regulation can only remove energy, exchange cannot change it, and only the task can add it. Without task input, the forecast therefore stays inside the bowl where it started, however far it runs and however often the phase switches. On the other hand, a cue adds deviation directly between steps, as treated in the next proposition. The proof is the discrete form of \cref{eq:Hdot}: the midpoint rule makes the step use one gradient $v$, so the exchange term vanishes exactly, and only the current step's matrices appear, which is why any phase sequence is allowed.

In words, regulation removes energy, exchange conserves it, and the task alone adds it. Without task input, the forecast therefore stays inside the bowl where it started, however far it runs and however often the phase switches. On the other hand, a cue adds deviation directly between steps, as treated next. The proof is the discrete form of \cref{eq:Hdot}: the midpoint rule makes the step use one gradient $v$, so the exchange term vanishes exactly, and only the current step's matrices appear, which is why any phase sequence is allowed.

Because every regulation value is positive, each step without task input also removes a positive share of the energy in every direction. We measure the distance from rest with the energy norm $\|z\|_K=\sqrt{z^\top Kz}=\sqrt{2H}$, built from the energy of \cref{eq:H}. In each step this norm shrinks by at least the factor $\rho=\max_\sigma\|C_\sigma\|_K<1$, the largest fraction of this distance that can remain after one step (\cref{lem:contraction} in \cref{app:proofs}).

\begin{proposition}[Bounded cue schedules]\label{prop:impulse}
Without task input, let a cue $d_h$ enter step $h$ as $z_{h+1}=C_\sigma(z_h+d_h)$, with $d_h=0$ between cues, and let cues satisfy $\|d_h\|_K\le\delta$, so that no cue gives a team at rest ($z=0$) more energy than $\tfrac12\delta^2$, and be at least $\tau$ steps apart. Then for every phase sequence and cue schedule,
\begin{equation}
\|z_h\|_K\le\rho^{h}\|z_0\|_K+\frac{\delta}{1-\rho^{\tau}}.\label{eq:cuebound}
\end{equation}
\end{proposition}
% In words, every step still shrinks whatever deviation is there by at least the factor $\rho$, but a cue first adds a kick of size at most $\delta$, and the step then acts on the kicked deviation, so the deviation can grow at a cue. Between cues the dynamics remove deviation, and the bound says that this balance stays below $\delta/(1-\rho^\tau)$, so spaced cues cannot pile up into a runaway prediction. The proof sums this fading where the $r$-th most recent cue contributes at most $\delta\rho^{r\tau}$, and the geometric series gives the bound on all cues together (full proof in \cref{app:proofs}). This is what a planner needs where any schedule it considers has a known worst case.

In words, every step still shrinks whatever deviation is there by at least the factor $\rho$, but a cue first adds a kick of size at most $\delta$, and the step then acts on the kicked deviation, so the deviation can grow at a cue. Between cues the dynamics remove deviation, and the bound says that this balance stays below $\delta/(1-\rho^\tau)$, so spaced cues cannot pile up into a runaway prediction. The proof sums this fading: the $r$-th most recent cue contributes at most $\delta\rho^{r\tau}$, and the geometric series gives the bound on all cues together (full proof in \cref{app:proofs}). This is what a planner needs where any schedule it considers has a known worst case. We express such bounds in \emph{cues' worth}, multiples of $\delta$. The bounds scale with $\delta$, so they hold whatever the size of a real cue turns out to be.

\vspace{-2mm}
\paragraph{When the resting point moves.} Both results describe one forecast with a fixed resting point $q^\star$. A live system recomputes $q^\star$ at every window, so cues that change behavior also move the bowl. \Cref{app:closed} extends the analysis to this case. A bound that holds for all time follows if the resting point leans more on other teams (a larger $w_{\mathrm{pop}}$ in \cref{eq:attractor}), which pins the bowl in place but makes forecasts less accurate, because the model then ignores more of the team's own behavior (\cref{prop:uniform} in \cref{app:results}). The energy norm lets a designer compute this trade-off between forecast accuracy and how far repeated cues can move the team before deployment.

\vspace{-2mm}
%\section{\model{} in the Loop}\label{sec:loop}
% \paragraph{The live loop.} In a live system the team is the plant, the headsets are the sensors and the prompts are the actuators, and \model{} is the model inside the controller. \Cref{alg:loop} shows how it runs, and \cref{fig:loop} in \cref{app:model} draws the same loop as a block diagram. All computation is closed form, so a forecast and a refit are cheap enough to run within the 16\,s stride (\cref{sec:use}).

\paragraph{The live loop.} In a live system the team is the plant, the headsets are the sensors and the prompts are the actuators, and \model{} is the model inside the controller. \Cref{alg:loop} shows how it runs, and \cref{fig:loop} in \cref{app:model} draws the same loop as a block diagram. All computation is closed form, so a forecast and a refit are cheap enough to run within the 16\,s stride.% (\cref{app:repro}).

\begin{algorithm}[t]
\caption{One pass of the live loop, every 16\,s.}
\label{alg:loop}
\small
\begin{algorithmic}[1]
\Require fitted parameters $\theta$, breakdown thresholds $\vartheta_j$, minimum cue spacing $\tau$
\State Read the new window $q_k$ and the remaining work $\pi_k$.
\State Update the three anchors and the resting point $q^\star$ by \cref{eq:attractor}.
\State $z_0\gets q_k-q^\star$; roll \cref{eq:step} forward to obtain $\hat q_{k+1},\dots,\hat q_{k+16}$.
\If{a team coordinate $j$ of $\hat q$ falls below $\vartheta_j$}
  \State Raise a warning.
\EndIf
\If{no cue was issued in the last $\tau$ windows \textbf{and} the forecast with a kick $d$ improves on the forecast without it by enough}
  \State Issue the cue $d$.
\EndIf
\end{algorithmic}
\end{algorithm}

\section{Evaluation Setup}\label{sec:eval}

We evaluate \model{} as a model of team dynamics through five questions. First, do real teams behave as the model assumes, drifting back toward a resting point after a deviation (\cref{sec:real})? Second, is each part of the model, and so each principle, needed to forecast teams it has not seen (\cref{sec:rq3})? Third, do the fitted parameters give a readable account of how teams work (\cref{sec:teams})? Fourth, is the model as accurate as larger models without its structure (\cref{sec:accuracy})? Fifth, are its forecasts useful in the closed loop, for early warning and for timing cues (\cref{sec:use})? The rest of this section explains our evaluation setup.
\vspace{-2mm}
\paragraph{Teams the model has not seen.} Every result is measured on teams left out of training. We use leave-one-team-out cross-validation. %: we fit on all other teams, forecast the one left out, and repeat for every team.
From every window $k\ge5$ of the held-out team, once five windows have been observed, each model forecasts the next 16 windows (256\,s). \model{} is trained with $h_{\mathrm{tr}}=8$ windows (128\,s), so the evaluation also tests forecasts beyond the training horizon.
\vspace{-2mm}
\paragraph{Error and comparison.} Error is the normalized mean squared error (\emph{nMSE}), scaled by the training variance so that 1 means an error as large as the data's natural variation. We report it separately for the 16 member and pair coordinates and for the four team coordinates. To compare two models, we compute each model's error on every held-out team and test whether one is consistently lower across the \nGroups{} teams with a paired Wilcoxon signed-rank test, which makes no assumption about how the errors are distributed. We fitted several variants of \model{}, which differ in how the matrices are built and in which optional terms they include (\cref{app:model}), and we report the variant with the lowest combined member-and-pair and team error.
\vspace{-2mm}
\paragraph{Baselines and controls.} To judge accuracy, we compare \model{} with simple rules and with learned models that have no built-in structure. The rules are persistence, the recent average and the session mean. The learned models are an iterated VAR(2), a direct multi-horizon ridge regression and a small MLP. They also train on \emph{symmetry-augmented} data, in which renaming the members gives 23 further valid copies, because \model{} has this symmetry built in. An \emph{attractor-only} control predicts the learned resting point without dynamics, which isolates what the dynamics add. % Early warning and the simulated cues are defined in \cref{sec:use}, where they are simulated and used.

\section{Results}\label{sec:results}

The results answer the three questions of \cref{sec:model} as follows, and \cref{tab:principles} summarizes them part by part.
\begin{itemize}[leftmargin=*, topsep=1pt]
\item \textbf{Where does the team rest?} At a point blended from the moment, the team and other teams, in proportions that differ by behavior. This is the model's strongest element.
\item \textbf{How does it return?} By sliding back without swinging, with recovery times of about 20--60\,s.
%\item \textbf{What moves it away?} The modelled channels, task progress and phase, have the expected direction but add no significant accuracy at this resolution, and the effect of cues remains to be measured in a deployment.
%\item \textbf{What moves it away?} Task progress and phase move the resting point and the recovery rate in the expected direction and carry the guarantees and the input channels, and a deployment can measure the effect of cues.
\item \textbf{What moves it away?} Task progress and phase, in the expected direction; a deployment will measure the effect of cues.

\end{itemize}

\subsection{Do real teams behave like a ball in a bowl?}\label{sec:real}

\paragraph{How we measure fading.} The guarantees describe the model, and whether the picture also holds for real teams is an empirical question. To test it directly, we ask how much of a team's deviation from its resting point survives over time. For every window $k$ of a held-out team, we compute $z_k=q_k-q^\star_k$ and the later deviation $z_{k,h}=q_{k+h}-q^\star_k$ from the same resting point, and we project one on the other:
\begin{equation}
m_h=\frac{\sum_k z_k^\top K z_{k,h}}{\sum_k z_k^\top K z_k}.\label{eq:memory}
\end{equation}
This $m_h$ is the share of a deviation that remains after $h$ windows. A value of 1 means that the deviation has not changed, 0 means that it is gone, and a negative value would mean that the team overshoots to the other side. Window noise averages out because the projection is signed. As a control test, we shuffle the order of the windows in each session and repeat the calculation. The shuffled session has the same states but no order in time, so its $m_h$ is close to zero, and any fading that real teams show beyond it comes from how they move through time.

\begin{figure}[t]
\centering
\includegraphics[width=\columnwidth]{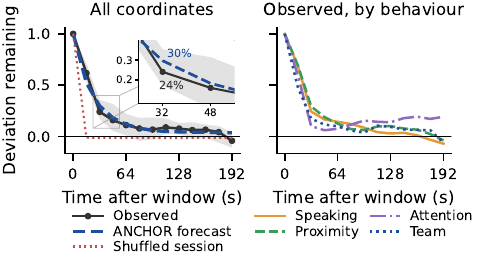}
\caption{Real deviations fade, on average close to the forecast. Share of a deviation that remains $h$ windows later, $m_h$, on held-out teams. Left: observed (band: interquartile range over teams), the \model{} forecast, and sessions with shuffled windows; the inset zooms on 24--56\,s, around the 32\,s values compared in the text. The 16\,s point shares half its window with the start (\cref{fig:windows} in \cref{app:data}), so comparisons start at 32\,s. Right: the same observed quantity for each behavior separately, computed from that behavior's coordinates alone (with $K$ replaced by the identity).}
\label{fig:memory}
\end{figure}

\vspace{-2mm}
% \paragraph{Real teams fade toward rest.} Real deviations fade the way a ball settles in a bowl, and on average close to the forecast (\cref{fig:memory}). After 32\,s, the first step free of window overlap, \memObsc\% of a deviation remains against a forecast of \memPredc\%. After 48\,s the values are \memObsd\% and \memPredd\%, after 64\,s \memObse\% and \memPrede\%, and after 80\,s \memObsf\% and \memPredf\%. Each further 16\,s removes about 30\% of what remains, and a constant fraction is what a ball settling in a bowl should show. Most of a deviation has faded after two minutes, which is why a training horizon of $h_{\mathrm{tr}}=8$ windows (128\,s) suffices. Shuffled sessions, whose resting points are computed in the same way, drop to about \memSurr\% at once, so the gradual fading of real teams comes from how they move through time rather than from how the resting point is computed.
\paragraph{Real teams fade toward rest.} Real deviations fade the way a ball settles in a bowl, and on average close to the forecast (\cref{fig:memory}). After 32\,s, the first step free of window overlap, \memObsc\% of a deviation remains against a forecast of \memPredc\%. Each further 16\,s removes about 30\% of what remains, and a constant fraction is what a ball settling in a bowl should show. Most of a deviation has faded after two minutes, which is why a training horizon of $h_{\mathrm{tr}}=8$ windows (128\,s) suffices. Shuffled sessions, whose resting points are computed in the same way, drop to about \memSurr\% at once, so the gradual fading of real teams comes from how they move through time rather than from how the resting point is computed.

% \paragraph{Reading the match with the forecast.} The match with the forecast is partly expected and should be read with care. The forecast comes from models fitted without each team and trained on the forecast error of \cref{eq:loss}, not on the $m_h$ of \cref{eq:memory}. Least-squares training of a linear relaxation does, however, tend to reproduce the average decay of its training data. The stronger tests are therefore the autoregressive control and the differences between behaviors. %, which follow.

\paragraph{Reading the match with the forecast.} The forecast comes from models fitted without each team and trained on the forecast error of \cref{eq:loss} rather than on the $m_h$ of \cref{eq:memory}. Because least-squares training of a linear relaxation tends to reproduce the average decay of its training data, the autoregressive control and the differences between behaviors, which follow, give the sharper tests.

\begin{table*}[t]
\caption{The hypothesis behind each part of the model and what held-out teams say about it. Each test replaces one part with the alternative shown and retrains the full model, which here also contains the sensed damping. Cells give the change in held-out nMSE, so a positive change means the part helps, followed by the number of teams for which \model{} is better and the paired Wilcoxon $p$. Errors are over 16--256\,s, except for P2 (16--48\,s, where dynamics act) and P3 (speaking balance). Shaded: supported ($p<0.05$). On par: the reduced model reaches the same accuracy, and these parts are kept because they carry a guarantee or the channel for task progress and cues. }
\label{tab:principles}
\small
\setlength{\tabcolsep}{3pt}
\begin{tabular}{>{\raggedright\arraybackslash}p{2.6cm}>{\raggedright\arraybackslash}p{4.3cm}>{\raggedright\arraybackslash}p{3.0cm}ccc}
\toprule
Part & Hypothesis & Alternative tested & Member and pair nMSE & Team nMSE & Verdict\\
\midrule
\rowcolor{ours}Resting point $q^\star$ (P1) & Teams return to a point set by recent behaviour, their own history and other teams & $q^\star$ from recent behaviour only & +19.3\% (17/17, 2e-5) & +16.5\% (14/17, 0.004) & \textbf{supported} \\
\rowcolor{ours} &  & $q^\star$ from session history only & +3.4\% (14/17, 0.035) & +9.7\% (14/17, 0.020) & \textbf{supported} \\
\rowcolor{ours}Regulation $R_\sigma$ (P2) & Deviations fade, so modelling the return helps at 16--48\,s & attractor only, no dynamics & +11.3\% (17/17, 2e-5) & +8.5\% (15/17, 5e-4) & \textbf{supported} \\
\rowcolor{ours}Team state $q$ (P3) & Team states are not reducible to members and pairs & computed from member forecast & -- & +270.3\% (16/17, 3e-5) & \textbf{supported} \\
\rowcolor{ours}Shared link gains in $K,R_\sigma,J_\sigma$ & Coordinates with the same role can share one value & unconstrained matrices & +14.4\% (16/17, 4e-4) & +9.7\% (17/17, 2e-5) & \textbf{supported} \\
Phase $\sigma$ (P4) & Teams regulate differently in exploration and consolidation & single phase & +0.1\% (9/17, 0.782) & -0.4\% (7/17, 0.611) & on par \\
Task input $g\,u$ (P5) & Task progress moves the resting point & no progress port & +1.6\% (9/17, 0.487) & +2.2\% (9/17, 0.644) & on par \\
Exchange $J_\sigma$ & Interaction passes between behaviours without loss & -- & \multicolumn{2}{c}{gains $|J_\sigma|\le\jMax$; regulation dominates} & small \\
Sensed damping $D(t)$ & Dominance and conflict add damping & no sensed damping & -0.0\% (8/17, 0.459) & -0.1\% (7/17, 0.431) & on par \\
Inertia $M$ & Trends persist across windows & momentum added & +0.2\% (11/17, 0.487) & +0.2\% (5/17, 0.579) & on par \\
Cue $d$ (P6) & Well-timed cues move a team & -- & \multicolumn{2}{c}{simulated in \cref{sec:use}} & simulated \\
 
\bottomrule
\end{tabular}
\end{table*}

\begin{itemize}[noitemsep, topsep=1pt, leftmargin=1em]
% \item{Against an autoregressive control:} Real teams also fade faster than a simple autoregressive process would. Shuffling is a weak control, since any autocorrelated process fades, so we also compare with an AR(1) surrogate fitted to each coordinate of each session. Measured the same way, the AR(1) surrogate keeps \arAllNull\% of a deviation after 32\,s, whereas real teams keep \arAllReal\%, and real teams keep less in all \arAllWins{} teams ($p=\arAllP$). At team level the values are \arTeamNull\% and \arTeamReal\% (real teams lower in \arTeamWins{} of \nGroups{}, $p=\arTeamP$). Real deviations therefore fade faster than autoregression predicts, partly because overlapping windows inflate its fitted lag. %Establishing a mechanism would need real perturbations.

\item{Against an autoregressive control.} Real teams also fade faster than a simple autoregressive process would. Because any autocorrelated process fades, we also compare with a stronger control, an AR(1) surrogate fitted to each coordinate of each session. Measured the same way, the AR(1) surrogate keeps \arAllNull\% of a deviation after 32\,s, whereas real teams keep \arAllReal\%, and real teams keep less in all \arAllWins{} teams ($p=\arAllP$). At team level the values are \arTeamNull\% and \arTeamReal\% (real teams lower in \arTeamWins{} of \nGroups{}, $p=\arTeamP$). Real deviations therefore fade faster than autoregression predicts, partly because overlapping windows inflate its fitted lag. %Delivered cues in a deployment would identify the mechanism behind this faster fading.

% \item{By behavior:} The match holds on average and at team level, but pair attention fades faster than forecast. It is closest for the team-level state, where 17\% remains at 32\,s against a forecast of 17\% ($p=0.96$). Speaking (\memSpkTwo\% against 35\%, $p=0.06$) and proximity (\memProxTwo\% against 39\%, $p=0.01$) fade somewhat faster at 32\,s and match from 48\,s on. Pair attention is the exception, since \memAttTwo\% remains against a forecast of 31\% ($p<0.001$) and the model overstates it up to 96\,s. Sparse hovers over virtual objects may partly explain this, since a noisy signal looks as if it fades at once. The result therefore holds for the team-level state and for the average, not for every behavior. It holds on the signed projection and not on raw energy, which noise keeps high in any one window (\cref{app:teams}).

\item{By behavior.} The match is closest for the team-level state, where 17\% remains at 32\,s against a forecast of 17\% ($p=0.96$). Speaking (\memSpkTwo\% against 35\%, $p=0.06$) and proximity (\memProxTwo\% against 39\%, $p=0.01$) fade slightly faster at 32\,s and match from 48\,s on. Pair attention fades fastest, with \memAttTwo\% remaining against a forecast of 31\% ($p<0.001$), so the model is conservative for attention up to 96\,s, consistent with brief, sparse hovers that make a signal look as if it fades at once. %The signed projection averages window noise out, which raw energy does not (\cref{app:teams}).

\end{itemize}

\subsection{Does each principle hold?}\label{sec:rq3}

% By a principle we mean one of the claims from team psychology that \model{} builds in (P1--P6 in \cref{sec:model}), such as that teams return to a resting point, each carried by one part of the model (\cref{tab:mapping}). A principle holds if removing its part makes forecasts of unseen teams worse. By this test, four parts are needed: the blended resting point, the dynamics, the team-level state and the shared link structure; the others make no detectable difference. \Cref{tab:principles} removes each element in turn and retrains, which is the central test of the paper, since a model that only fits the data would not have to survive it. To read the table, take the first row: when the resting point uses recent behavior only, error rises by 19\% for members and pairs, \model{} is better for all 17 teams, and the difference is significant.

By a principle we mean one of the claims from team psychology that \model{} builds in (P1--P6 in \cref{sec:model}), each carried by one part of the model (\cref{tab:mapping}). A principle holds if removing its part makes forecasts of unseen teams worse. By this test, four parts are needed: the blended resting point, the dynamics, the team-level state and the shared link structure; the others are on par and carry the guarantees and input channels. \Cref{tab:principles} removes each element in turn and retrains, which is the central test of the paper because it checks every part on teams the model has not seen. To read the table, take the first row: when the resting point uses recent behavior only, error rises by 19\% for members and pairs, \model{} is better for all 17 teams, and the difference is significant.

The table shows that the resting point must blend its anchors and that dynamics add to it. Team states must be modelled at the team level. A given member is silent \fracSilentMember\% of the time, so a forecast of expected speaking times looks evenly shared even when one member dominates, and balance derived from the member forecast has \hierBalRatio{} times the error of forecasting it directly (\hierBalDerived{} against \hierBalExplicit), which is what multilevel team theory predicts~\cite{kozlowski2000}. The results for a resting point from recent behavior only, the dynamics, speaking balance and the shared structure also hold after correcting for the 17 tests of the table (\cref{app:significance}).

% The other elements answer the third question in a simple way. Phases, the task port, sensed damping and inertia do not change accuracy significantly, and the exchange gains are small ($|J_\sigma|\le\jMax$), so at this resolution teams slide back to rest rather than swing. The phase also has little chance to matter, because it changes once in \phOnce{} of \nGroups{} sessions, typically \phFirstMin{} minutes in, so only \phSoon\% of forecasts meet a switch within 48\,s (\cref{app:teams}). We keep the phases and the port because they carry the guarantees and the channel for the task and for cues, and the selected model drops the sensed damping.

The other elements answer the third question simply. Phases, the task port, sensed damping and inertia are on par with the reduced model, and the exchange gains stay within $|J_\sigma|\le\jMax$, so at this resolution teams slide back to rest rather than swing. The phase changes once in \phOnce{} of \nGroups{} sessions, typically \phFirstMin{} minutes in, so \phSoon\% of forecasts meet a switch within 48\,s (\cref{app:teams}).%, and longer tasks with more switches would test it further. 
We keep the phases and the port because they carry the guarantees and the channel for the task and for cues, and the selected model leaves out the sensed damping.

\subsection{What does the fitted model say about teams?}\label{sec:teams}

\begin{figure}[t]
\centering
\includegraphics[width=\columnwidth]{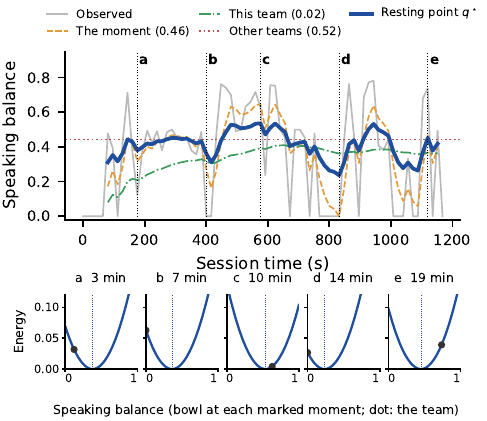}
\caption{The bowl moves. Speaking balance of held-out team \bowlTeam{} (the longest session), with the resting point $q^\star$ recomputed every window as the blend of its three anchors (fitted weights in parentheses). Bottom: this coordinate's bowl, $\tfrac12K_{ii}(x-q^\star)^2$, at five marked moments that include the lowest and highest resting point, and where the team actually was (dot). The bottom slides with the conversation while the team moves up and down its walls.}
\label{fig:bowl}
\end{figure}

\begin{figure}[t]
\centering
\includegraphics[width=\columnwidth]{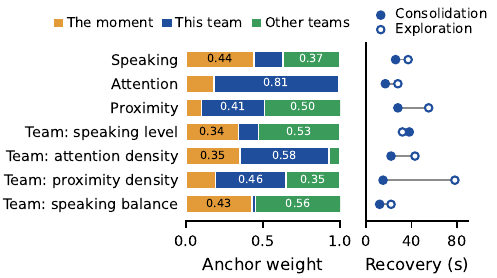}
\caption{Reading a team off the fitted model. Left: what each behavior's resting point follows, from the learned anchor weights. Right: recovery time constants in each phase. \Cref{tab:structure} in \cref{app:teams}  lists all values.}
\label{fig:weights}
\end{figure}

Each behavior rests on a different anchor: shared attention on the team's own history, proximity on habit and norms, and speaking balance on the moment. The fitted model therefore reads like a description of how teams work (\cref{fig:weights}). \emph{Who attends together} is a trait of the team (session weight \wSessionAtt), while \emph{how close members stand} mixes the team's own habit with other teams (weights \wSessionProx{} and \wPopProx). \emph{How evenly they talk} belongs to the moment: the resting balance follows the moment (\wTeamRecentSpeakingBalance) and other teams (\wTeamPopSpeakingBalance), with almost no weight on the team's own history (\wTeamSessionSpeakingBalance), and it recovers fastest, in \tauConsBal--\tauExplBal\,s. \Cref{fig:bowl} shows this on one held-out team: over the session the resting balance moves between \bowlLo{} and \bowlHi, following the moment but pulled back toward other teams (\bowlPop). Recovery is fast overall, with time constants in consolidation of \tauConsAtt\,s for shared attention, \tauConsConv\,s for speaking and \tauConsProx\,s for proximity, and it is slower in exploration. As images are decided, the fitted port moves the resting point toward more shared attention and more co-presence, and resting balance rises by about 0.25 on its 0--1 scale from start to finish, a descriptive reading of the fitted port.

\subsection{How accurate is \model{}?}\label{sec:accuracy}

\begin{table}[t]
%\caption{Forecast error of each model relative to \model{} on held-out teams: how much higher its error is, averaged over teams (\%; positive: higher error than \model). Bold: significant (paired Wilcoxon over the \nGroups{} teams, $p<0.05$). Verdict for \model: better if significantly better in at least one column and worse in none; on par if no column is significant. Params: values fitted from the data (\model: \bestParams). The anchored models use \model{}'s fitted anchors, which favours them. Win counts, $p$-values and absolute errors are in \cref{tab:relative_detail,tab:absolute} in \cref{app:accuracy}. }
\caption{Forecast error of each model relative to \model{} on held-out teams: how much higher its error is, averaged over teams (\%; positive: higher error than \model). Bold: significant (paired Wilcoxon over the \nGroups{} teams, $p<0.05$). Verdict for \model: better if significantly better in at least one column and worse in none; on par if no column is significant. The anchored models use \model{}'s fitted anchors, which favors them. Win counts, $p$-values and absolute errors are in \cref{tab:relative_detail,tab:absolute} in \cref{app:accuracy}.}
\label{tab:baselines}
\small
\setlength{\tabcolsep}{2pt}
\begin{tabular}{lrccccc}
\toprule
& & \multicolumn{2}{c}{Member and pair} & \multicolumn{2}{c}{Team} & \\
\cmidrule(lr){3-4}\cmidrule(lr){5-6}
Model & Params & 16--48\,s & 16--256\,s & 16--48\,s & 16--256\,s & Verdict\\
\midrule
Attractor only & $21$ & \textbf{+11.3} & +1.5 & \textbf{+8.5} & +0.1 & better \\
Persistence & $0$ & \textbf{+38.8} & \textbf{+49.2} & \textbf{+32.1} & \textbf{+43.5} & better \\
Recent average & $0$ & \textbf{+23.1} & \textbf{+22.6} & \textbf{+16.2} & \textbf{+22.1} & better \\
Session mean & $0$ & \textbf{+28.0} & \textbf{+13.1} & \textbf{+46.4} & \textbf{+36.4} & better \\
VAR(2) & $820$ & \textbf{+15.3} & +10.4 & \textbf{+16.6} & +11.2 & better \\
VAR(2) + aug. & $820$ & \textbf{+6.4} & +8.3 & \textbf{+7.6} & +9.0 & better \\
Direct ridge & $43{,}840$ & \textbf{+28.0} & \textbf{+15.3} & \textbf{+23.2} & \textbf{+14.2} & better \\
Ridge + aug. & $43{,}840$ & -1.1 & +1.5 & -1.1 & +1.4 & on par \\
MLP & $14{,}944$ & \textbf{+39.3} & \textbf{+14.7} & \textbf{+44.9} & \textbf{+25.9} & better \\
MLP + aug. & $14{,}944$ & \textbf{+33.9} & +10.0 & \textbf{+32.3} & +15.2 & better \\
GRU & $27{,}168$ & \textbf{+38.7} & \textbf{+15.7} & \textbf{+36.8} & +15.7 & better \\
GRU + aug. & $27{,}168$ & \textbf{+28.9} & \textbf{+7.3} & \textbf{+26.1} & +8.5 & better \\
Anchored scalar & $28$ & \textbf{+1.7} & +0.2 & +0.0 & +0.4 & better \\
Anchored free & $434$ & \textbf{+1.8} & +0.7 & \textbf{+4.4} & +1.9 & better \\
Anchored free + aug. & $434$ & \textbf{+1.5} & +0.2 & +0.1 & +0.8 & better \\
 
\bottomrule
\end{tabular}
\end{table}

\model{} performs on par with augmented ridge, with \bestParams{} parameters against about 40,000 (\cref{tab:baselines}). Its 16--256\,s error is the lowest at both levels, with ridge at \relRidgeAugDyL\% and \relRidgeAugTmL\% higher error ($p=\relRidgeAugDyLP$ and $\relRidgeAugTmLP$), so the two are on par. Persistence, which assumes that the team stays as it is now, has about half again \model{}'s error over 16--256\,s (\relPersDyL\% for members and pairs, \relPersTmL\% for the team). The other learned baselines, including a GRU sequence model, have significantly higher error over 16--48\,s. Error falls in clear steps as parts are added (member and pair level, 16--48\,s), from \relEmaDyS\% higher error than \model{} for the recent average, with nothing fitted, to \relStaticDyS\% for the resting point alone (21 values) and \relScalarDyS\% when each behavior decays toward it at its own rate (seven more). \model{}'s remaining 24 values, for stiffness, links, phases and exchange, close this last gap ($p=\relScalarDySP$), and at team level the two tie (\relScalarTmS\%). With the same anchors, 434 free values have \relFreeDyS\% and \relFreeTmS\% higher error.

\begin{figure}[t]
\centering
\includegraphics[width=\columnwidth]{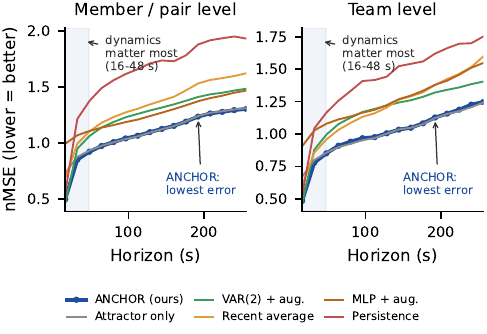}
\caption{Forecast error (nMSE, lower is better) by horizon on held-out teams, against the iterated and simple baselines (augmented ridge is in \cref{tab:baselines}), for the 16 member and pair coordinates (left) and the four team coordinates (right). Grey: attractor-only control; shaded: the first 16--48\,s. Values above 1 at long horizons still beat assuming an average team (\cref{app:accuracy}).}
\label{fig:horizon}
\end{figure}

\Cref{fig:horizon} shows the error by horizon. Over 16--48\,s, \model{} also beats the attractor-only control, which has \relStaticDyS\% and \relStaticTmS\% higher error ($p=\relStaticDySP$ and $\relStaticTmSP$), whereas the two meet far ahead because the forecast is then simply where the team relaxes to. The dynamics therefore supply the path from the present to the resting point, which is the horizon at which a system would act. \model{} stays significantly better than assuming an average team up to \dyUsefulHolmSec\,s (pairs) and \tmUsefulHolmSec\,s (team) after correcting for the 16 horizons tested (\cref{app:significance}). Because the structure is small, \model{} learns from few teams: on mean error it stays below augmented ridge at every training-set size, significantly up to eight teams at member and pair level and four at team level (\cref{fig:de} in \cref{app:accuracy}). With very few teams, the resting point alone gives a robust forecast.

% \begin{figure}[t]
% \centering
% \includegraphics[width=\columnwidth]{figs/data_efficiency.pdf}
% \caption{Error versus the number of training teams, as the median over held-out teams and random training subsets. With very few teams the resting point alone is the safest forecast.}
% \label{fig:de}
% \end{figure}

\subsection{Can forecasts give early warning and cue timing?}\label{sec:use}

% \begin{figure}[t]
% \centering
% \includegraphics[width=\columnwidth]{figs/rollout_energy.pdf}
% \caption{The guarantee in action. Left: a 24-window forecast for a held-out team (thin: observed, thick: forecast) relaxes smoothly from the current state to the resting point. Right: the energy $H_k$ decays and never exceeds its initial value plus the energy supplied by the progress input (dashed), as \cref{prop:passivity} guarantees.}
% \label{fig:rollout}
% \end{figure}

\paragraph{Guarantees on the fitted model.} On the fitted model, a cue's predicted effect fades within about a minute, and repeated cues stay within a few cues' worth. The fitted contraction factor is $\rho=\rhoContraction$, so the predicted effect of one cue halves within \rhoHalfLifeSec\,s even in the slowest direction, team proximity density during exploration, and faster for speaking (\cref{fig:attractor}). With one cue every 32\,s, the bound of \cref{prop:impulse} is \rhoGainTwo{} cues' worth. A single cue therefore gives a short nudge and does not change where the team rests, unless the team's own later behaviour moves the anchors. In 60-window (16\,min) rollouts for all held-out teams, with phase switching and task input, the energy balance held at \passiveFrac\% of steps, and every forecast coordinate stayed between $-\longMax$ and $\longMax$, so the forecasts stayed bounded even 16\,min ahead; \cref{fig:rollout} in \cref{app:teams} shows one such forecast. %VAR(2) was also stable on these data, but a VAR can only be checked after each fit, whereas these results hold for any parameter values and so survive retraining.
In the live loop, where the resting point is recomputed, the computed bound is about \clCertSession{} cues' worth over 16\,min and grows slowly with time (\cref{app:closed}). This yields a design insight that shared attention has almost no population weight (\clWpopAtt), so sustained cues on attention gradually become part of what the model treats as normal. %On one laptop thread a 16-step forecast takes \timeForecast\,ms and fitting on 16 teams takes \timeFit\,s, well within the 16\,s stride.

\vspace{-1mm}
\paragraph{Early warning.} \model{} anticipates conversation collapse well, mostly because it knows where each team rests. A \emph{breakdown} is a team coordinate falling below the 25th percentile of the training teams: conversation collapse for speaking balance (below about 0.2, close to one effective speaker), loss of shared focus for attention density, and low proximity for proximity density. A warning is a forecast that a breakdown absent now will begin within the next \ewWindowSec\,s. At this threshold, \model{} warns of conversation collapse with AUROC \ewAucConv, catching \ewDetConv\% of onsets at a \ewFar\% false-alarm rate, and of loss of shared focus with AUROC \ewAucAtt. At the 15th and 35th percentiles collapse AUROC is about 0.75, and \model{} stays within 0.02 of the best model (\cref{tab:sensitivity} in \cref{app:warning}). The attractor-only control ties it on both, so most of this signal comes from knowing where the team rests; plausibly, teams whose resting balance sits near the threshold are the ones that collapse. For low proximity, \model{} reaches \ewAucProx{} and augmented ridge \ewBestOtherProx. \Cref{tab:warning} gives all three. For chronically uneven teams, an alarm level calibrated per team would reduce extra alarms (\cref{fig:case} in \cref{app:warning}). At about one false alarm per 80\,s of healthy time, which suits a facilitator's display.

\begin{table}[t]
\caption{Warning of a breakdown that starts within the next \ewWindowSec\,s and is absent now, on held-out teams. AUC: AUROC (0.5 = chance); Det.: share of onsets detected when alarms fire at 20\% of origins without a breakdown; Warn: mean warning time (s). Shaded: \model; bold: best in each column.}
\label{tab:warning}
\footnotesize
\setlength{\tabcolsep}{2.2pt}
\begin{tabular}{lccccccccc}
\toprule
& \multicolumn{3}{c}{Conversation collapse} & \multicolumn{3}{c}{Loss of shared focus} & \multicolumn{3}{c}{Low proximity}\\
\cmidrule(lr){2-4}\cmidrule(lr){5-7}\cmidrule(lr){8-10}
Model & AUC & Det. & Warn & AUC & Det. & Warn & AUC & Det. & Warn\\
\midrule
\rowcolor{ours}ANCHOR (ours) & \textbf{0.85} & \textbf{74\%} & 31 & \textbf{0.76} & 57\% & 35 & 0.66 & 34\% & 29 \\
Attractor only & \textbf{0.85} & 73\% & 31 & \textbf{0.76} & 56\% & 35 & 0.62 & 28\% & 31 \\
Recent average & 0.83 & 68\% & 32 & \textbf{0.76} & \textbf{58\%} & 37 & 0.63 & 26\% & 32 \\
VAR(2) + aug. & 0.80 & 64\% & 30 & 0.69 & 43\% & 36 & 0.65 & 39\% & 28 \\
Ridge + aug. & 0.80 & 60\% & 31 & 0.71 & 42\% & 35 & \textbf{0.68} & \textbf{43\%} & 32 \\
MLP + aug. & 0.74 & 50\% & 31 & 0.72 & 51\% & 37 & 0.56 & 25\% & 33 \\
 
\bottomrule
\end{tabular}
\end{table}

\begin{figure}[t]
\centering
\includegraphics[width=\columnwidth]{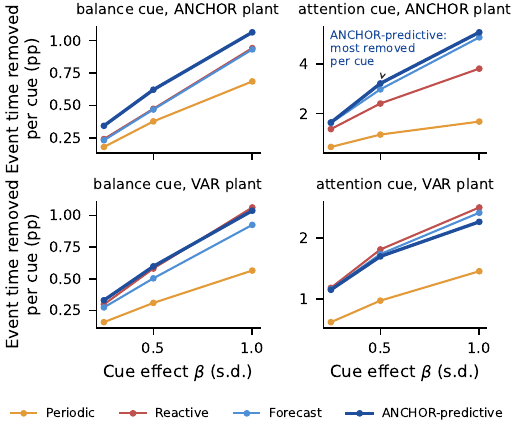}
\caption{Simulated cues under an assumed effect $\beta$. Breakdown time removed per cue (percentage points of the episode; higher is better), on \model{} itself (top) and on an independently learned VAR plant (bottom). \model-predictive (thick blue) is on top in \ivMpcWinsPh{} of 6 settings on its own plant and in \ivMpcWinsVar{} of 6 on the VAR plant.}
\label{fig:intervention}
\end{figure}

\vspace{-1mm}
\paragraph{Timing cues, in simulation.} If cues work, timing them by forecast gets more out of each one than cueing on a schedule. The recorded sessions contain no cues, so we simulate cues whose effect we assume and sweep, on two plants: \model{} itself and an independently learned VAR that the controller has never seen. Each episode starts from the first six windows of a held-out team and runs for 40 windows on the plant, which steps forward from its own forecast plus an error drawn from its one-step errors on real data. A cue adds an assumed $\beta$ training standard deviations to speaking balance or to shared attention in the window it is issued, and all policies keep at least two windows between cues (protocol in \cref{app:cues}). We compare four policies. Periodic cueing issues a cue every eight windows, whatever the team does. Reactive cueing issues a cue once the breakdown is visible in the current window. Forecast cueing issues a cue when the lowest forecast over the next 48\,s falls below a level calibrated on training teams. \emph{\model-predictive} compares the forecast with and without a cue, and it issues a cue when the predicted drop in the coming shortfall exceeds such a level.
At an assumed effect of 0.5 training standard deviations, it removes \ivEffMpc{} points of breakdown time per cue, which is \ivEffRatioMpcPeriodic{} times as much as periodic cueing (\ivEffPeriodic) and \ivEffRatioMpcReactive{} times as much as reactive cueing (\ivEffReactive), pooled over plants (\cref{fig:intervention}). It is the most efficient policy in \ivMpcWinsPh{} of 6 settings on its own plant and in \ivMpcWinsVar{} of 6 on the VAR plant, where reactive cueing is competitive. Selectivity trades total effect for efficiency: for balance cues on \model{}'s plant, from 13.3\% of time in breakdown, periodic cueing reaches 11.4\% with 4.7 cues per 10\,min and \model-predictive reaches 12.1\% with 1.8 (\cref{tab:intervention} in \cref{app:cues}). These results show how well a policy times its cues for a given cue effect, and a deployment will measure that effect.

% {\color{red} ------ Salma stopped here ---- }

\section{Related Work}\label{sec:related}

\paragraph{Human-in-the-loop CPS.} Schirner et al.~\cite{schirner2013} frame systems that sense people and act with them as human-in-the-loop CPS, in which the human is part of the loop rather than an external user. Prior systems adapt to one person, for example by timing collision warnings to a driver~\cite{elmalaki2018sentio} or by planning around a driver's latent trust~\cite{sheng2022trust}, and recent work keeps decisions fair across the several people a system serves~\cite{zhao2024fairo,zhao2024fina} and spreads privacy risk evenly across the people whose data train it~\cite{zhao2026finp}. We study a team as the plant, where the state lies in the relations among people and the actuator is a prompt that they may or may not follow.

\vspace{-1mm}
\paragraph{Sensing and predicting teams.} Wearable sensing showed that turn-taking and proximity relate to team outcomes~\cite{pentland2012,zhang2018teamsense}, and smartphones can infer who converses with whom~\cite{luo2013socialweaver}. In collaborative MR, MURMR~\cite{murmr} supplies the sensing pipeline and data we use, S2S~\cite{s2s} relates sensed signals to perceived collaboration, and MoCoMR~\cite{mocomr} simulates individual behavior. TeamLLM~\cite{teamllm} predicts the next window with language models, but it degrades sharply when iterated. Classical social-influence models let opinions relax toward a weighted consensus anchored by individual prejudices~\cite{degroot1974,friedkin1990,proskurnikov2017}, and dyadic models let partners relax toward a set point~\cite{gottman2002}. \model{} shares their spirit, and it extends the idea to multimodal interaction at three levels, with learned anchors and formal guarantees.

\vspace{-1mm}
\paragraph{Physics-informed learning and passivity.} Hamiltonian and dissipative networks use energy structure as a prior for physical systems~\cite{greydanus2019,zhong2020}, and port-Hamiltonian networks add dissipation and external forcing~\cite{desai2021}. Other work builds stability into deep dynamics by learning a Lyapunov function with the model~\cite{manek2019}, or verifies neural Lyapunov and barrier certificates formally~\cite{abate2021fossil}. Passivity and dissipativity are also established tools for designing CPS~\cite{antsaklis2013}, including switched systems under quantization~\cite{zhu2012passivity}, and we carry the same reasoning to a human plant. Teams have no known energy law, so here each part of the structure is tested on held-out data (\cref{tab:principles}). A shared energy is the classical route to stability under arbitrary switching~\cite{liberzon2003}, and the \pH{} form builds it in~\cite{vanderschaft2014}.

\section{Discussions and Conclusion}\label{sec:discussion}\label{sec:conclusion}

% \paragraph{Why port-Hamiltonian?} At this resolution the data select a very simple member of the family: exchange is small and regulation dominates, and a scalar relaxation toward the same resting point nearly ties \model{} (\cref{sec:accuracy}). This does not count against the form, because such a relaxation, with decay between 0 and 1, is itself a \pH{} system with $J_\sigma=0$ and diagonal $K$ and $R_\sigma$. What the form adds is the family itself. Its guarantees hold for every parameter value, whereas a free linear map, although stable in all \rtFits{} retrainings on 2--16 teams (radius at most \rtFreeMax), has to be rechecked after each fit. It names each part after a claim about teams, which turns model design into a set of experiments. Richer effects such as exchange and inertia also fit inside the same energy balance, so they can be tested without losing the guarantees.
\paragraph{Why port-Hamiltonian?} At this resolution the data select a simple member of the family, in which regulation dominates exchange, and a scalar relaxation toward the same resting point comes close to \model{} (\cref{sec:accuracy}). Such a relaxation, with decay between 0 and 1, is itself a \pH{} system with $J_\sigma=0$ and diagonal $K$ and $R_\sigma$, so the data choose within the family. What the form adds is the family itself. Its guarantees hold for every parameter value, whereas a free linear map, which was stable in all \rtFits{} retrainings on 2--16 teams (radius at most \rtFreeMax), is checked after each fit. It names each part after a claim about teams, which turns model design into a set of experiments. Richer effects such as exchange and inertia also fit inside the same energy balance, so they can be tested without losing the guarantees.

\paragraph{What the picture says.} Over the minutes-long course of a session, team interaction is well described by relaxation toward an anchored resting point. This gives ideas from team psychology a computable form, in which equilibria, self-regulation and emergent states become numbers that can be estimated and tested. We offer the energy picture as a description of what teams do on average, not as a claim that they literally are \pH{} systems.

\paragraph{Scope.} The study covers one task, one room size and four-person peer teams, so the fitted anchors and gains describe this task, and the contribution is the structure, which carries to new tasks by refitting. The configuration was selected on the evaluation folds, and several neighboring configurations are within 1\% of it.

%\section{Conclusion}\label{sec:conclusion}

% \model{} models a team as a ball in an energy bowl, and it turns that picture into a port-Hamiltonian plant with \bestParams{} readable parameters and guarantees that hold by construction. On held-out teams, deviations fade close to the forecast on average and at the team level, and the anchors, the relaxation and team-level states each earn their place. The next step is to close the loop with real cues in a deployment study.

% \paragraph{Conclusion.} \model{} turns the ball-in-a-bowl picture into a port-Hamiltonian plant with \bestParams{} readable parameters and guarantees that hold by construction. It describes what teams do on average, in a form whose guarantees hold for every fit. The next step is to close the loop with real cues in a deployment study.

\paragraph{Conclusion.} \model{} turns the ball-in-a-bowl picture into a port-Hamiltonian plant with \bestParams{} readable parameters and guarantees that hold by construction. For design, a system should act on the team's trajectory and spend cues on conversation balance, where breakdowns are short and recurrent, while attention calls for sustained support. The next step is to close the loop with real cues in a deployment study.

\begin{acks}
This work was supported by the National Science Foundation under grant number 2339266.
\end{acks}

%\newpage

\bibliographystyle{ACM-Reference-Format}
\bibliography{references}

\clearpage
\appendix
% Appendices of anchor_paper.tex; these do not count toward the page limit.

\section*{Appendices}
The appendices give supporting detail. \Cref{app:notation} lists the notation, \cref{app:data} describes data preprocessing and \cref{app:model} gives the complete model. \Cref{app:proofs} contains full proofs, \cref{app:closed} treats the case in which the resting point is recomputed, \cref{app:results} reports additional results on significance, accuracy, the fitted model, early warning and simulated cues. % and \cref{app:repro} explains how to reproduce them.

% -----------------------------------------------------------------------------
\section{Notation}\label{app:notation}
% -----------------------------------------------------------------------------

\Cref{tab:notation} lists every symbol of the main text.

\begin{table}[h]
\caption{All symbols used in the main text. Each is also introduced where it first appears.}
\label{tab:notation}
\small
\setlength{\tabcolsep}{3pt}
\begin{tabular}{>{\raggedright\arraybackslash}p{2.2cm}>{\raggedright\arraybackslash}p{5.8cm}}
\toprule
Symbol & Meaning\\
\midrule
$k$, $h$ & window at which a forecast starts; steps ahead (16\,s each)\\
$q_k$, $\hat q_{k+h}$ & team state (20 numbers); its forecast\\
$q^\star$ & resting point (equilibrium, or attractor)\\
$\bar q_{\mathrm{rec}},\bar q_{\mathrm{ses}},\bar q_{\mathrm{pop}}$ & recent, session and population anchors\\
$w_{\mathrm{rec}},w_{\mathrm{ses}},w_{\mathrm{pop}}$ & learned anchor weights\\
$z=q-q^\star$ & deviation from the resting point\\
$H$, $K$ & energy; stiffness (steepness of the bowl)\\
$\kappa$, $\gamma_\ell$, $L_\ell$ & log-steepness, one per behaviour and one per team coordinate; gain and Laplacian of link $\ell$\\
$\chi$, $L_{\mathrm{cons}}$ & gain and matrix tying the team coordinates to the member and pair coordinates\\
$\mathbf 1_n$ & vector of $n$ ones\\
$R_\sigma$, $J_\sigma$ & regulation (removes energy); exchange (moves energy)\\
$\zeta_\sigma$ & log-recovery rate, one per behaviour and one per team coordinate\\
$j_{\sigma,SA}$, $j_{\sigma,AP}$; $S_{SA}$, $S_{AP}$ & exchange gains; their skew-symmetric link matrices\\
$\sigma$ & task phase (exploration or consolidation)\\
$M$, $D(t)$ & inertia; sensed damping (tested variants, not selected)\\
$\pi_k$, $u_h$ & remaining work; forecast change in it\\
$g=Kb$ & port gain, with one learned value of $b$ per behaviour and one per team coordinate\\
$A_\sigma$, $C_\sigma$, $G_\sigma$, $\bar u_h$ & $(J_\sigma-R_\sigma)K$; one-step maps of the forecast; input at the step midpoint\\
$\theta$, $\nu$, $h_{\mathrm{tr}}$ & all parameters; inverse training variances in the loss; training horizon\\
$d_h$, $\delta$, $\tau$ & cue; bound on its size; minimum cue spacing\\
$\|z\|_K$, $\rho$ & energy norm $\sqrt{2H}$; shrink factor per step\\
$m_h$ & share of a deviation remaining after $h$ windows\\
$\beta$ & assumed cue effect in simulation (training standard deviations)\\
\bottomrule
\end{tabular}
\end{table}

% -----------------------------------------------------------------------------
\section{Data and Preprocessing}\label{app:data}
% -----------------------------------------------------------------------------

\begin{figure}[h]
\centering
\begin{tikzpicture}[font=\scriptsize, x=0.085cm, y=0.45cm]
\foreach \k/\c/\t in {0/blue!35/{window $k$}, 1/orange!45/{window $k{+}1$}, 2/green!45/{window $k{+}2$}} {
  \fill[\c] (16*\k,-\k) rectangle (16*\k+32,-\k+0.55);
  \draw[black!50, line width=0.3pt] (16*\k,-\k) rectangle (16*\k+32,-\k+0.55);
  \node[anchor=east] at (-1,-\k+0.27) {\t};
}
\fill[black!22] (16,-1) rectangle (32,-0.45);
\draw[black!50, line width=0.3pt] (16,-1) rectangle (32,-0.45);
\node[font=\tiny] at (24,-0.73) {shared};
\draw[densely dashed, black!50] (32,0.8) -- (32,-2.5);
\node[anchor=west, font=\tiny] at (49,-0.73) {$h{=}1$: 16\,s shared};
\node[anchor=west, font=\tiny] at (65,-1.73) {$h{=}2$: none shared};
\draw[->] (0,-2.6) -- (84,-2.6);
\foreach \x in {0,16,32,48,64} { \draw (\x,-2.52) -- (\x,-2.68) node[below=-1pt] {\x}; }
\node[anchor=north east] at (84,-2.65) {time (s)};
\end{tikzpicture}
\caption{Windows are 32\,s long and start every 16\,s. Window $k{+}1$ shares its first 16\,s with window $k$ (grey), so a comparison one step ahead partly compares the same recording with itself. Window $k{+}2$ starts where window $k$ ends, so 32\,s is the first step free of overlap.}
\label{fig:windows}
\end{figure}
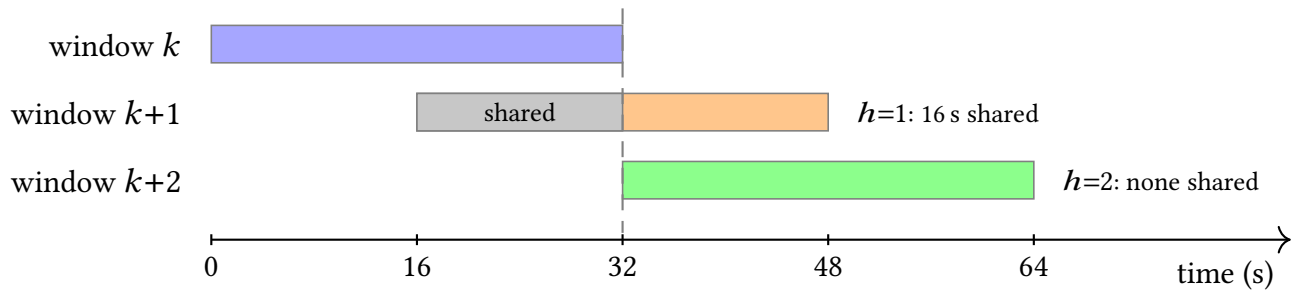

\paragraph{Windows.} We use the windowed values of the MURMR dataset~\cite{murmr}: 32\,s windows with a 16\,s stride, consecutive within each session, indexed by $k=0,1,\dots$. Every coordinate is normalized to $[0,1]$. \Cref{fig:windows} shows how consecutive windows overlap.

 The speaking time of member $i$ is divided by 32\,s, and since the conversation graph is broadcast, every outgoing edge of $i$ carries this value.

\paragraph{Attention and proximity.} The attention of pair $ij$ is the total time both members' gaze hovers simultaneously on the same image, $\sum\big(\min(e_i,e_j)-\max(s_i,s_j)\big)$ over overlapping hover intervals on one image, ignoring overlaps below 0.013\,s, divided by 32\,s. Hovers are events on virtual objects and are not verified fixations. The proximity of pair $ij$ is the number of seconds in the window with distance at most 0.4572 on the logged $X$--$Z$ floor plane, divided by 33, the number of whole-second stamps in a window. Logged units are treated as meters. The split of the proximity anchor between session and population shifts with preprocessing choices more than the other weights, so we read it as a mix of the two.

\paragraph{Team coordinates.} From each window we compute $q^t=(\bar s,\bar a,\bar r,\eta)$: the mean speaking time $\bar s$ over the four members, the mean joint-attention weight $\bar a$ and mean proximity weight $\bar r$ over the six pairs, and the speaking balance $\eta=-\sum_i\omega_i\log\omega_i/\log4$ with speaking shares $\omega_i=s_i/\sum_js_j$ ($\eta=0$ when nobody speaks).

\paragraph{Anchors.} With $n_{\mathrm{rec}}=5$ windows and decay $\lambda_{\mathrm{rec}}=0.7$, the recent anchor is the exponential average
\[
\bar q_{\mathrm{rec},k}=\textstyle\sum_{j=0}^{n_{\mathrm{rec}}-1}\lambda_{\mathrm{rec}}^jq_{k-j}\Big/\sum_{j=0}^{n_{\mathrm{rec}}-1}\lambda_{\mathrm{rec}}^j;
\] the session anchor is $\bar q_{\mathrm{ses},k}=\frac{1}{k+1}\sum_{j\le k}q_j$; the population anchor $\bar q_{\mathrm{pop}}$ is the mean over all windows of the training teams.

\paragraph{Sensed damping (tested, not selected).} Speaking dominance $\phi_i=|s^{\mathrm{rec}}_i/\sum_js^{\mathrm{rec}}_j-\tfrac14|$ uses the recent anchor $s^{\mathrm{rec}}$ of the speaking times. Overrides come from the object-interaction logs: an override $i\to j$ is recorded when member $i$ places an image whose last placement was by $j\ne i$ with a different label. The count $o_{ij}$ is the exponential average of these events over $n_{\mathrm{rec}}$ windows, and pair $ij$ receives the conflict $o_{ij}+o_{ji}$.

\paragraph{Progress signals.} We compared three task-progress signals. (i) The label-based error is the fraction of images whose current label differs from the group's \emph{final} label, which uses end-of-session information and is therefore not causal. (ii) Coverage is the fraction of images never placed. (iii) Remaining work, used throughout the paper, is the fraction of images that have not been placed with a label that then stayed unchanged for four windows (64\,s). All three relate to shared attention in the same way, with within-team Spearman correlations between attention density and the signal of \rhoAttEps{} (label-based), \rhoAttCov{} (coverage) and \rhoAttStab{} (remaining work), so the causal signal loses nothing. At a forecast origin $k$ we extrapolate the signal as $\hat\pi_{k+h}=\mathrm{clip}(\pi_k-\upsilon h,0,1)$ with rate $\upsilon=\max(\upsilon_{\mathrm{pop}},\,\text{exponential average of }\pi_{j-1}-\pi_j)$, where $\upsilon_{\mathrm{pop}}$ is the mean decrease per window in the training teams while $\pi>0$. The recent level $\bar\pi_k$ is the exponential average of $\pi$ over the last $n_{\mathrm{rec}}$ windows.

\paragraph{The 1\,s interaction table.} MURMR also releases per-second pairwise features, whose speaking indicators we used only for the timescale diagnostic of \cref{app:teams}. Each member's per-second speaking indicator is identical across the three pairs that contain the member. The per-second timelines end earlier than the windowed ones, so the two resolutions are not sample-aligned. Aggregated into 32\,s windows and aligned at the best offset for each team, the indicator correlates with the windowed conversation weights at $r=0.58$--$0.92$ per team.

% -----------------------------------------------------------------------------
\section{Complete Model Specification}\label{app:model}
% -----------------------------------------------------------------------------

The selected model is fully specified by \cref{eq:Kd,eq:R,eq:J}, the port $g=Kb$ of \cref{sec:moves} and \cref{tab:params}. Here we add what the main text leaves out. We write the coordinates in four blocks: S (speakers, 4), A (attention pairs, 6), P (proximity pairs, 6) and T (team coordinates, 4).

\paragraph{A circuit analogy.} \Cref{tab:analogy} shows the same port-Hamiltonian structure in an electrical circuit and in a team, for readers who know the former.

\begin{table}[h]
\caption{The same structure in a circuit and in a team. The team column is a modelling hypothesis, not a physical law. The symbols are defined in \cref{sec:return,sec:moves} and listed in \cref{tab:notation} in \cref{app:notation}.}
\label{tab:analogy}
\small
\setlength{\tabcolsep}{3pt}
\begin{tabular}{>{\raggedright\arraybackslash}p{1.6cm}>{\raggedright\arraybackslash}p{2.7cm}>{\raggedright\arraybackslash}p{3.3cm}}
\toprule
Term & Circuit & Team\\
\midrule
Energy $H$ & Charge stored in capacitors & Distance from the usual pattern\\
Regulation $R_\sigma$ & Resistors, which dissipate & Self-regulation back to the pattern\\
Exchange $J_\sigma$ & Lossless coupling & Talk turning into shared focus\\
Port $g\,u$ & Voltage source & Task progress shifting the pattern\\
Kick $d$ & Current pulse & A prompt to the team\\
Phase $\sigma$ & Switch position & Exploration or consolidation\\
\bottomrule
\end{tabular}
\end{table}

\paragraph{Tie between levels.} $L_{\mathrm{cons}}=\sum_{B\in\{S,A,P\}}v_Bv_B^\top$ with $v_B=e_{T_B}-\frac{1}{n_B}\sum_{c\in B}e_c$, where $e_c$ is the unit vector of coordinate $c$, $T_B$ is the team coordinate of behaviour $B$ and $n_B$ is its number of coordinates. Then $\tfrac12\chi z^\top L_{\mathrm{cons}}z$ is the sum of the terms $\tfrac12\chi(t-m)^2$ of \cref{sec:return}. The fitted value is $\chi=\chiFit$.

\paragraph{Sensed damping (tested, not selected).} With sensed damping, regulation becomes $R_\sigma+D(t)$ with
\[
D(t)=\mathrm{diag}\big(\mu_{\mathrm{dom}}\phi,\ \mu_{\mathrm{ov}}\mathbf o,\ \mu_{\mathrm{ov}}\mathbf o,\ \mathbf 0\big),
\]
where $\phi\in\mathbb{R}^4$ stacks the speaking dominances, $\mathbf o\in\mathbb{R}^6$ the pair conflicts $o_{ij}+o_{ji}$ (\cref{app:data}), and $\mu_{\mathrm{dom}},\mu_{\mathrm{ov}}\ge0$ are learned. The selected model has $D=0$.

\paragraph{Anchor weights.} The weights of \cref{eq:attractor} are softmaxes of free logits, one triple per behaviour and one per team coordinate.

\paragraph{Permutation-equivariant family.} Any two coordinates of the member and pair block have one of 15 symmetric relation types---SS (same, different); AA, PP and AP (same pair, pairs sharing a member, disjoint pairs); SA and SP (member in pair, member not in pair)---and the 7 cross-type relations (SA, SP, AP) admit an orientation, giving 7 skew-symmetric basis matrices. The member and pair block of $K$ is $B^2+\epsilon I$ and $R=F^2$ ($\epsilon=10^{-3}$), with $B,F$ linear combinations of the symmetric basis and $J$ a linear combination of the skew basis. Same-type skew couplings (for example member $i$ to member $j$) would require an orientation between interchangeable members and are therefore excluded by symmetry.

\paragraph{Unconstrained family.} The member and pair block of $K$ is $BB^\top+\epsilon I$, $R=FF^\top$ with lower-triangular $B,F$, and $J=W-W^\top$ with free $W$.

\paragraph{Second-order variant (inertia).} This variant adds a momentum $p$ with diagonal inertia $M$, so the state is $(z,p)$ and the energy Hessian is $\mathrm{blkdiag}(K,M^{-1})$, with $J=\begin{bmatrix}0&I\\-I&J^{pp}_\sigma\end{bmatrix}$, $R=\mathrm{blkdiag}(0,R^{pp}_\sigma+D(t))$ and $g=(0,Kb)$, with $M$ diagonal per modality and team coordinate; the initial momentum is $p_0=M\,\overline{\Delta q}$, where $\overline{\Delta q}$ is the exponential average of recent window-to-window differences.

\paragraph{Observer (+obs).} Over the last six windows, the observer alternates the frozen step with the update $\hat q\leftarrow\hat q^-+\ell\odot m_k\odot(q_k-\hat q^-)$, where the gain $\ell\in(0,1)$ is learned per modality and $m_k$ marks the observed modalities.

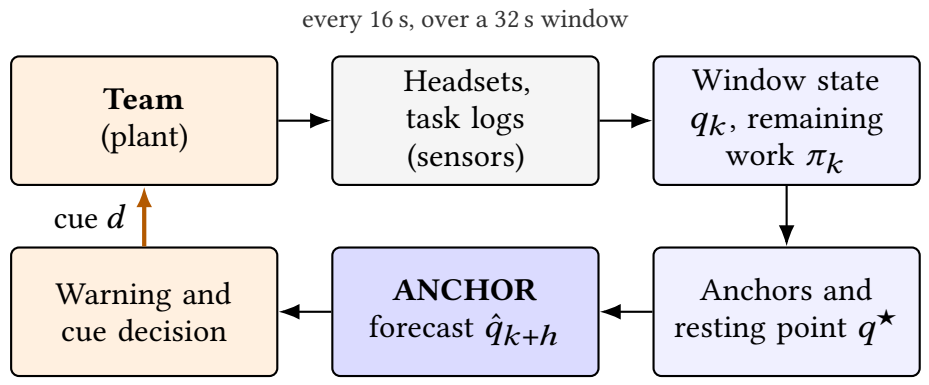
\begin{figure}[h]
\centering
\begin{tikzpicture}[font=\scriptsize, >={Latex[length=1.4mm]}, node distance=4mm and 3.5mm,
  box/.style={draw, rounded corners=1.5pt, align=center, inner sep=2pt, minimum height=0.85cm, text width=1.62cm, fill=#1}, box/.default=white]
\node[box=orange!12] (team) {\textbf{Team}\\(plant)};
\node[box=gray!8, right=of team] (sense) {Headsets, task logs\\(sensors)};
\node[box=blue!6, right=of sense] (state) {Window state $q_k$, remaining work $\pi_k$};
\node[box=blue!6, below=of state] (anch) {Anchors and resting point $q^\star$};
\node[box=blue!14, left=of anch] (fore) {\textbf{\model{}} forecast $\hat q_{k+h}$};
\node[box=orange!12, left=of fore] (dec) {Warning and cue decision};
\draw[->] (team) -- (sense); \draw[->] (sense) -- (state); \draw[->] (state) -- (anch); \draw[->] (anch) -- (fore); \draw[->] (fore) -- (dec);
\draw[->, thick, orange!70!black] (dec) -- node[left, align=right, text=black] {cue $d$} (team);
\node[above=1pt of sense, font=\tiny, gray!40!black] {every 16\,s, over a 32\,s window};
\end{tikzpicture}
\caption{The closed loop. The team is the plant, and \model{} is the model inside the controller. In this paper the cue arrow is simulated, since the recorded sessions contain no cues.}
\label{fig:loop}
\end{figure}

\paragraph{Live loop.} \Cref{fig:loop} draws the loop of \cref{alg:loop} as a block diagram.

\paragraph{Learning and selection.} The loss is
\[
\mathcal L(\theta)=\frac{1}{20N}\sum_{\text{origins }k}\sum_{h=1}^{h_{\mathrm{tr}}}\sum_{c=1}^{20}\nu_c\big(\hat q_{k+h,c}-q_{k+h,c}\big)^2+\lambda_{\mathrm{reg}}\|\theta\|^2,
\]
with $\theta$ all parameters, $N$ the number of origin--step pairs with an observed target, $\nu_c$ the inverse training variance of the coordinate's modality (or team coordinate), $\lambda_{\mathrm{reg}}=10^{-6}$ ($10^{-4}$ for the unconstrained family), minimized by L-BFGS-B (at most 300 iterations) with exact gradients. The configuration is selected by the sum of pooled member and pair- and team-level nMSE over 16--256\,s. \Cref{tab:hyper} lists all hyperparameters.

\begin{table}[h]
\caption{Hyperparameters.}
\label{tab:hyper}
\small
\begin{tabular}{p{3.6cm}p{4.2cm}}
\toprule
Recent anchor $n_{\mathrm{rec}}$, decay $\lambda_{\mathrm{rec}}$ & 5 windows, 0.7\\
Training horizon $h_{\mathrm{tr}}$ / evaluation horizon & 8 / 16 windows\\
Phase threshold on $\hat\pi$ & 0.5\\
Stability window for $\pi$ & 4 windows (64\,s)\\
Optimizer & L-BFGS-B, $\le$300 it., exact gradients\\
Regularization $\lambda_{\mathrm{reg}}$ & $10^{-6}$ ($10^{-4}$ unconstrained)\\
VAR(2) ridge / direct ridge & 1 / 10\\
MLP & 32 tanh units, weight decay $10^{-3}$, 400 it.\\
Symmetry augmentation & 23 relabellings (MLP: 7)\\
Breakdown threshold $\vartheta_j$ & 25th percentile of training teams\\
Alarm window / false-alarm rate & 4 windows / 20\%\\
Simulation & 40 windows, 30 seeds, cue spacing $\tau=2$\\
\bottomrule
\end{tabular}
\end{table}

\paragraph{Baselines.} VAR(2) is $q_{k+1}=W[q_k;q_{k-1};1]$, fitted by ridge regression and iterated. Direct ridge uses one linear map per horizon whose inputs are the last five windows (100 values), the session mean (20), the extrapolated progress inputs (16) and a constant, which gives 137 features and about $4\times10^4$ parameters. The MLP uses the same standardized inputs and one hidden layer. Augmentation adds the 23 non-identity relabellings of each training team, since relabelling the members permutes speakers and pairs and leaves team coordinates unchanged.

% -----------------------------------------------------------------------------
\section{Proofs}\label{app:proofs}
% -----------------------------------------------------------------------------

We give proofs for the first-order model. They use only that the energy is quadratic with a positive definite Hessian, so they also hold for the second-order variant with $K$ replaced by $\mathrm{blkdiag}(K,M^{-1})$, except that its regulation is singular and only $\|C_\sigma\|_K\le1$ is guaranteed. For a matrix $X$, we write $\|X\|_K=\max_{z\ne0}\|Xz\|_K/\|z\|_K=\|K^{1/2}XK^{-1/2}\|_2$.

\begin{lemma}[Contraction]\label{lem:contraction}
Let $K\succ0$, $J=-J^\top$ and $R\succeq0$, and let $A=(J-R)K$. Then $I-\tfrac12A$ is invertible and $C=(I-\tfrac12A)^{-1}(I+\tfrac12A)$ satisfies $\|C\|_K\le1$. If $R\succ0$, then $\|C\|_K<1$.
\end{lemma}
\begin{proof}
With $\xi=K^{1/2}z$, the map becomes $\tilde C=(I-\tfrac12\tilde A)^{-1}(I+\tfrac12\tilde A)$ with $\tilde A=K^{1/2}(J-R)K^{1/2}=\tilde J-\tilde R$, where $\tilde J$ is skew-symmetric and $\tilde R\succeq0$. If $(I-\tfrac12\tilde A)v=0$ for $v\ne0$, then $\|v\|^2=\tfrac12v^\top\tilde Av=-\tfrac12v^\top\tilde Rv\le0$, a contradiction, so the matrix is invertible. For any $v$, let $\psi=(I-\tfrac12\tilde A)^{-1}v$. Then
\begin{align*}
\|\tilde Cv\|^2-\|v\|^2&=\|(I+\tfrac12\tilde A)\psi\|^2-\|(I-\tfrac12\tilde A)\psi\|^2\\
&=\psi^\top(\tilde A+\tilde A^\top)\psi=-2\psi^\top\tilde R\psi\le0,
\end{align*}
with strict inequality for $v\ne0$ when $\tilde R\succ0$. By compactness of the unit sphere, $\|\tilde C\|_2<1$ in that case.
\end{proof}

In the first-order model every diagonal entry of $R_\sigma$ is an exponential and any sensed damping is non-negative, so $R_\sigma\succ0$ and $\rho=\max\|C_\sigma\|_K<1$. The reported value $\rho=\rhoContraction$ is this maximum over all origins and phases of the fitted model.

\paragraph{Proof of \cref{prop:passivity}.} With $H_h=\tfrac12z_h^\top Kz_h$ and $K$ symmetric, $H_{h+1}-H_h=\tfrac12(z_{h+1}-z_h)^\top K(z_{h+1}+z_h)=v^\top(z_{h+1}-z_h)$ with $v=\tfrac12K(z_h+z_{h+1})$, which is the energy gradient at the midpoint. The implicit midpoint rule gives $z_{h+1}-z_h=(J_\sigma-R_\sigma)v+g\bar u_h$. Hence $H_{h+1}-H_h=v^\top J_\sigma v-v^\top R_\sigma v+v^\top g\bar u_h$. The first term is zero because $J_\sigma$ is skew-symmetric, and the second is non-positive because $R_\sigma\succeq0$. The argument uses only the matrices of the current step, so it holds for every phase sequence. \qed

\paragraph{Proof of \cref{prop:impulse}.} Set $d_h=0$ at steps without a cue, so that $z_{h+1}=C_{\sigma_h}(z_h+d_h)$. Unrolling gives
\[
z_h=\Phi(h,0)z_0+\sum_{h_i<h}\Phi(h,h_i)d_{h_i},\qquad \Phi(h,j)=C_{\sigma_{h-1}}\cdots C_{\sigma_j},
\]
and $\|\Phi(h,j)\|_K\le\rho^{h-j}$. Order the cues before $h$ from the most recent, $r=0,1,\dots$. Because they are at least $\tau$ steps apart, the $r$-th satisfies $h-h_i\ge r\tau$. Therefore $\|z_h\|_K\le\rho^h\|z_0\|_K+\delta\sum_{r\ge0}\rho^{r\tau}=\rho^h\|z_0\|_K+\delta/(1-\rho^\tau)$. \qed

\begin{lemma}[Shift of the resting point]\label{lem:shift}
For a fixed phase and a constant input $u$, the first-order model has the unique equilibrium $z^\ast=\Gamma u$ with $\Gamma=-A_\sigma^{-1}g$, which is also the fixed point of \cref{eq:step}. This $\Gamma$ is the ``Shift'' of \cref{tab:structure}, averaged over origins and phases.
\end{lemma}
\begin{proof}
$A_\sigma$ is invertible because $K\succ0$ and the symmetric part of $J_\sigma-R_\sigma$ is negative definite. The continuous equilibrium solves $A_\sigma z^\ast+gu=0$. The discrete fixed point solves $(I-\tfrac12A_\sigma)z^\ast=(I+\tfrac12A_\sigma)z^\ast+gu$, which is the same equation.
\end{proof}

\begin{proposition}[Invariance to relabelling the members]\label{prop:invariance}
Let $P$ be the $20\times20$ permutation induced by renaming the four members, which permutes speakers and pairs and leaves team coordinates fixed. Then $PKP^\top=K$, $PJ_\sigma P^\top=J_\sigma$ and the baseline regulation commutes with $P$, and sensed damping, when used, permutes with the data. If the population anchor is symmetric, $P\bar q_{\mathrm{pop}}=\bar q_{\mathrm{pop}}$, then renaming the input data renames the forecast in the same way.
\end{proposition}
\begin{proof}
The per-behaviour diagonal terms are constant on each block. Each link matrix is defined by a relation between coordinates, such as ``speaker $i$ belongs to pair $ij$'', that renaming preserves, so these matrices commute with $P$. Sensed damping, recent anchors and session anchors are computed per member and per pair, so they permute with the data. The attractor weights are shared within each kind of behaviour, so every quantity in the forecast is either invariant or permutes with $P$, and the solution of a permuted linear system is the permuted solution.
\end{proof}

The reported results use the per-coordinate population mean, which is not exactly symmetric, for example because member labels are tied to headsets. Replacing it with its symmetric version changes held-out error by \symDyChg\% at the member and pair level ($p=\symDyP$) and by \symTmChg\% at the team level ($p=\symTmP$).

% -----------------------------------------------------------------------------
\section{When the Resting Point Is Recomputed}\label{app:closed}
% -----------------------------------------------------------------------------

\Cref{prop:passivity,prop:impulse} hold for one forecast. A live system recomputes $q^\star_k$ at every window, so the energies of successive forecasts are measured from different points and do not add up. We state what still holds, for the first-order model.

\paragraph{Real teams.} All coordinates of $q$ lie in $[0,1]$. Let $D$ be the largest energy distance between two points of $[0,1]^{20}$, $\bar\rho=\sup\|C_\sigma\|_K$ and $\bar\gamma=\sup\|G_\sigma\|_K$. Let $W_{\mathrm{rec}}$ and $W_{\mathrm{ses}}$ be the diagonal matrices of anchor weights, and let $S=\sum_{j=0}^4\lambda_{\mathrm{rec}}^j$.

\begin{proposition}[Forecasts of a real team]\label{prop:measured}
For any observations in $[0,1]^{20}$, the resting point $q^\star_k$ stays in $[0,1]^{20}$. For every window $k$, phase sequence, input with $|\bar u_h|\le1$ and cue schedule as in \cref{prop:impulse},
\begin{equation*}
\|\hat q_{k+h}-q^\star_k\|_K\le\bar\rho^{\,h}D+\frac{\delta}{1-\bar\rho^{\,\tau}}+\frac{\bar\gamma}{1-\bar\rho}.
\end{equation*}
Once five windows have been observed, the resting point moves by at most
\begin{equation*}
\|q^\star_{k+1}-q^\star_k\|_K\le D\Big(\frac{\|W_{\mathrm{rec}}\|_K}{S}+\frac{\|W_{\mathrm{ses}}\|_K}{k+2}\Big).
\end{equation*}
\end{proposition}
\begin{proof}
The anchors are averages of points in the box, and \cref{eq:attractor} is a convex combination in each coordinate, so $q^\star_k$ is in the box and $\|z_0\|_K\le D$. Within a forecast $q^\star_k$ is fixed, so unrolling $z_{h+1}=C_\sigma(z_h+d_h)+G_\sigma\bar u_h$ as in \cref{prop:impulse} gives the first bound. For the second, the session anchor changes by $(q_{k+1}-\bar q_{\mathrm{ses},k})/(k+2)$, whose norm is at most $D/(k+2)$. The recent anchor changes by $(q_{k+1}-v)/S$, where $v=\sum_{j=0}^3(\lambda_{\mathrm{rec}}^j-\lambda_{\mathrm{rec}}^{j+1})q_{k-j}+\lambda_{\mathrm{rec}}^4q_{k-4}$ is a convex combination of observations, so its norm is at most $D/S$. The population anchor is constant.
\end{proof}

For the fitted model, $D=\clDiam$ and $\bar\gamma/(1-\bar\rho)=\clPort$. These bounds need no assumption about how teams respond to cues, and they are conservative, because they allow the resting point to travel anywhere in the box.

\paragraph{The model feeding on itself.} To ask how far repeated cues could move the resting point, let the model's own states build the anchors. Write $e_h$ for the state minus $\bar q_{\mathrm{pop}}$, let $W_{\mathrm{pop}}=I-W_{\mathrm{rec}}-W_{\mathrm{ses}}$, and take no task input:
\begin{equation}
e_{h+1}=a_h+C_\sigma(e_h+d_h-a_h),\quad a_h=W_{\mathrm{rec}}\bar e_{\mathrm{rec},h}+W_{\mathrm{ses}}\bar e_{\mathrm{ses},h},\label{eq:selffed}
\end{equation}
where $\bar e_{\mathrm{rec},h}$ and $\bar e_{\mathrm{ses},h}$ are the anchors of $e$. Only the population anchor pulls toward a fixed point, since the other two follow the state, so that the bowl moves with the ball.

\begin{proposition}[Cues in the self-fed loop]\label{prop:selffed}
Fix the phase sequence and let $e_0=0$.
(i) \cref{eq:selffed} is linear in the history, so $e_T=\sum_h\Psi(T,h)d_h$, and for every schedule with $\|d_h\|_K\le\delta$ and spacing at least $\tau$, $\|e_T\|_K\le\delta B_\tau(T)$ with $B_\tau(T)=\max_{\text{schedules}}\sum_h\|\Psi(T,h)\|_K$.
(ii) If $W_{\mathrm{pop}}\succ0$ and a constant cue $\bar d$ is applied in a fixed phase, the only constant solution is $e^\ast=W_{\mathrm{pop}}^{-1}(I-C_\sigma)^{-1}C_\sigma\bar d$.
(iii) If $W_{\mathrm{pop}}=0$, every constant vector solves \cref{eq:selffed} without cues, so $e=0$ is not asymptotically stable.
\end{proposition}
\begin{proof}
(i) The anchors are linear in past states with coefficients that depend only on the step, so superposition gives the sum and the triangle inequality gives the bound. Dynamic programming computes the maximum over schedules exactly. (ii) At a constant solution both anchors equal $e^\ast$, so $e^\ast-a=W_{\mathrm{pop}}e^\ast$, and \cref{eq:selffed} reads $(I-C_\sigma)W_{\mathrm{pop}}e^\ast=C_\sigma\bar d$, where $I-C_\sigma$ is invertible by \cref{lem:contraction}. (iii) If the whole history equals $c$, then $a_h=c$ and $e_{h+1}=c$.
\end{proof}

A simple small-gain argument for a time-uniform bound would require $\|C\|_K+\|(I-C)W_{\mathrm{rec}}\|_K+\|(I-C)W_{\mathrm{ses}}\|_K<1$, and for the model fitted on all teams this sum is \clKappa. We therefore compute $B_\tau(T)$ directly, in both phases and for every time of one phase switch. With one cue per 32\,s, $B_2(T)$ is \clCertShort{} over 8\,min, \clCertSession{} over 16\,min and \clCertLong{} over 32\,min, against \rhoGainTwo{} for a fixed resting point. With one cue per 16\,s it is \clCertEvery{} over 16\,min, against \rhoGainOne. A simulation in exploration with cues on one kind of behaviour at a time gives \clSimSpk{} cues' worth for speaking, \clSimProx{} for proximity and \clSimAtt{} for shared attention after 16\,min.

The fitted population weight explains the difference. It is \wPopConv{} for speaking and \wPopProx{} for proximity, but only \clWpopAtt{} for shared attention, which is therefore close to case (iii). The constant solution in (ii) would lie about \clSteady{} cues' worth away in the worst direction at one cue per 32\,s, but the approach is extremely slow, because the session anchor averages all past windows: the simulated deviation is \clSimAttLong{} after 2000 windows, about nine hours. Simulation confirms the approach: with weights $(0.3,0.3,0.4)$ the simulated state is within \clFormulaErr\% of it after 4000 windows.

\begin{proposition}[Time-uniform bound]\label{prop:uniform}
In the self-fed loop of \cref{eq:selffed} with cues $\|d_h\|_K\le\delta$, let $\kappa=\sup_\sigma\big(\|C_\sigma\|_K+\|(I-C_\sigma)W_{\mathrm{rec}}\|_K+\|(I-C_\sigma)W_{\mathrm{ses}}\|_K\big)$. If $\kappa<1$, then $\|e_h\|_K\le\rho\,\delta/(1-\kappa)$ for all $h$.
\end{proposition}
\begin{proof}
From \cref{eq:selffed}, $e_{h+1}=C_\sigma(e_h+d_h)+(I-C_\sigma)(W_{\mathrm{rec}}\bar e_{\mathrm{rec},h}+W_{\mathrm{ses}}\bar e_{\mathrm{ses},h})$. The anchors are convex combinations of past states, so if all past states have norm at most $M$, then $\|e_{h+1}\|_K\le\kappa M+\rho\delta$. Taking $M=\rho\delta/(1-\kappa)$ closes the induction from $e_0=0$.
\end{proof}

Over the leave-one-team-out fits $\kappa$ reaches \flZeroKappa, so the condition calls for a larger population weight. A large population weight in effect fixes the bowl in place and trades some accuracy for it. Blending the anchor weights toward the population with share $\lambda$, without refitting, gives $\kappa=\flEightKappa$ at $\lambda=0.8$ and $\kappa=\flNineKappa$ at $\lambda=0.9$, with bounds of \flEightBound{} and \flNineBound{} cues' worth for all time. Member and pair nMSE then rises from \flZeroDy{} to \flEightDy{} and \flNineDy{}, and team nMSE from \flZeroTm{} to \flEightTm{} and \flNineTm. Refitting the other parameters under the floor is a natural way to reduce this cost.

\Cref{prop:selffed} describes the model feeding on itself, which is how a planner sees the loop, and the computed bound covers one phase switch, as in \phOnce{} of the \nGroups{} sessions.

% -----------------------------------------------------------------------------
\section{Additional Results}\label{app:results}
% -----------------------------------------------------------------------------

\subsection{Significance and multiple tests}\label{app:significance}

Every comparison of two models uses the error of each on the same \nGroups{} held-out teams. A paired Wilcoxon signed-rank test gives a $p$-value, the probability of a difference at least this consistent if the two models were equally good, and we call a result significant when $p<0.05$. Each test then has a 5\% chance of a false positive, so a table of many tests is likely to contain one by chance; with 17 independent tests the chance of at least one is about 58\%. We therefore also apply the Holm correction~\cite{holm1979}, which sorts the $m$ $p$-values and compares the smallest with $0.05/m$, the next with $0.05/(m-1)$, and so on, stopping at the first that fails. This keeps the chance of any false positive among all $m$ tests below 5\%. In \cref{tab:principles} ($m=17$), the results for a resting point from recent behaviour only, the dynamics, speaking balance and the shared structure pass, and a resting point from session history only is significant before the correction ($p=\sesOnlyDyP$ and $\sesOnlyTmP$). For the horizon comparison of \cref{fig:skill} ($m=16$), the vertical lines mark how far \model{} stays better after the correction.

\subsection{Forecast accuracy}\label{app:accuracy}

This part gives the detail behind \cref{sec:accuracy}: per-cell comparisons, absolute errors, errors by horizon, team and quantity, robustness to sensor corruption, data efficiency and the anchored baselines.

\begin{table}[t]
\caption{Robustness (member and pair nMSE, 16--256\,s; lower is better). Shift: trained on clean data, tested on corrupted inputs. Matched: trained and tested on corrupted inputs. +obs: \model{} with a model-based observer. Clean data: \model{} \rbCleanPh, VAR \rbCleanVar, MLP \rbCleanMlp.}
\label{tab:robust}
\footnotesize
\setlength{\tabcolsep}{2.4pt}
\begin{tabular}{lcccccccc}
\toprule
& \multicolumn{4}{c}{Shift} & \multicolumn{4}{c}{Matched}\\
\cmidrule(lr){2-5}\cmidrule(lr){6-9}
Corruption & +obs & \model & VAR & MLP & +obs & \model & VAR & MLP\\
\midrule
Noise $\sigma=0.05$ & 1.14 & 1.13 & 1.13 & 1.21 & 1.05 & 1.05 & 1.31 & 1.18 \\
Noise $\sigma=0.1$ & 1.47 & 1.47 & 1.17 & 1.29 & 1.11 & 1.11 & 1.38 & 1.20 \\
Noise $\sigma=0.2$ & 2.99 & 2.97 & 1.29 & 1.46 & 1.14 & 1.14 & 1.38 & 1.23 \\
Dropout $p=0.1$ & 1.01 & 1.01 & 1.12 & 1.16 & 1.01 & 1.01 & 1.12 & 1.16 \\
Dropout $p=0.3$ & 1.03 & 1.03 & 1.13 & 1.17 & 1.04 & 1.03 & 1.13 & 1.17 \\
Dropout $p=0.5$ & 1.04 & 1.04 & 1.14 & 1.17 & 1.06 & 1.04 & 1.15 & 1.18 \\
 
\bottomrule
\end{tabular}
\end{table}

\begin{table}[t]
\caption{All model families and ablations (nMSE on held-out teams, lower is better). Params: free parameters, counted as in \cref{tab:params}. The selected model is within 1\% of several neighbours. The rows from ``attractor: recent only'' on remove one part from the switched topology-tied model with sensed damping (54 parameters), as in \cref{tab:principles}; diag, equiv and full denote the topology-tied, looser-tie and free-matrix families, friction is the sensed damping, and scale coupling is the tie $\chi$ between levels.}
\label{tab:ablation}
\small
\setlength{\tabcolsep}{2.6pt}
\begin{tabular}{lcccc}
\toprule
& & \multicolumn{2}{c}{Member and pair} & Team\\
\cmidrule(lr){3-4}
Variant & Params & 16--48\,s & 16--256\,s & 16--256\,s\\
\midrule
Selected model & 52 & 0.758 & 1.082 & 1.002 \\
1st-order, diag & 45 & 0.762 & 1.088 & 1.011 \\
1st-order, diag, switched & 54 & 0.760 & 1.083 & 1.002 \\
1st-order, equiv & 69 & 0.769 & 1.093 & 1.013 \\
1st-order, equiv, switched & 95 & 0.771 & 1.086 & 0.998 \\
1st-order, full & 424 & 0.850 & 1.168 & 1.075 \\
1st-order, full, switched & 684 & 0.852 & 1.239 & 1.079 \\
2nd-order, diag & 52 & 0.759 & 1.086 & 1.009 \\
2nd-order, diag, switched & 61 & 0.761 & 1.088 & 1.016 \\
2nd-order, equiv & 76 & 0.763 & 1.088 & 1.011 \\
2nd-order, equiv, switched & 102 & 0.768 & 1.092 & 1.015 \\
2nd-order, full & 431 & 0.822 & 1.128 & 1.033 \\
2nd-order, full, switched & 691 & 0.864 & 1.175 & 1.062 \\
attractor: recent only & 40 & 0.887 & 1.293 & 1.152 \\
attractor: session only & 40 & 0.768 & 1.099 & 1.063 \\
no task-progress port & 47 & 0.771 & 1.123 & 1.057 \\
no friction features & 52 & 0.758 & 1.082 & 1.002 \\
no scale coupling & 53 & 0.759 & 1.082 & 1.001 \\
attractor only, no dynamics & 21 & 0.848 & 1.099 & 1.009 \\
 
\bottomrule
\end{tabular}
\end{table}

\begin{table}[t]
\caption{Team-level error by quantity (nMSE, 16--256\,s; lower is better). Attention density is the hardest quantity for every model.}
\label{tab:team}
\small
\setlength{\tabcolsep}{3pt}
\begin{tabular}{lcccc}
\toprule
Model & Speaking & Attention & Proximity & Speaking\\
& level & density & density & balance\\
\midrule
ANCHOR (ours) & 0.86 & 1.19 & 1.13 & 0.82 \\
Attractor only & 0.88 & 1.20 & 1.14 & 0.82 \\
Persistence & 1.22 & 1.61 & 1.58 & 1.25 \\
Recent average & 1.07 & 1.38 & 1.35 & 1.02 \\
VAR(2) + aug. & 0.97 & 1.61 & 1.25 & 0.86 \\
MLP + aug. & 1.11 & 1.61 & 1.26 & 1.04 \\
 
\bottomrule
\end{tabular}
\end{table}

\begin{table*}[t]
\caption{Forecast error of each model relative to \model{} on held-out teams, in detail: change in error averaged over teams (\%; positive: higher error than \model), number of teams for which \model{} is better, and paired Wilcoxon $p$. \Cref{tab:baselines} summarizes these values.}
\label{tab:relative_detail}
\small
\setlength{\tabcolsep}{4pt}
\begin{tabular}{lcccc}
\toprule
& \multicolumn{2}{c}{Member and pair} & \multicolumn{2}{c}{Team}\\
\cmidrule(lr){2-3}\cmidrule(lr){4-5}
Model & 16--48\,s & 16--256\,s & 16--48\,s & 16--256\,s\\
\midrule
Attractor only & +11.3\% (17/17, 2e-5) & +1.5\% (12/17, 0.057) & +8.5\% (15/17, 5e-4) & +0.1\% (9/17, 0.747) \\
Persistence & +38.8\% (17/17, 2e-5) & +49.2\% (17/17, 2e-5) & +32.1\% (16/17, 3e-5) & +43.5\% (17/17, 2e-5) \\
Recent average & +23.1\% (17/17, 2e-5) & +22.6\% (16/17, 3e-5) & +16.2\% (16/17, 3e-4) & +22.1\% (14/17, 0.001) \\
Session mean & +28.0\% (17/17, 2e-5) & +13.1\% (12/17, 0.006) & +46.4\% (16/17, 5e-5) & +36.4\% (14/17, 0.002) \\
VAR(2) & +15.3\% (16/17, 3e-5) & +10.4\% (12/17, 0.080) & +16.6\% (14/17, 5e-4) & +11.2\% (11/17, 0.071) \\
VAR(2) + aug. & +6.4\% (12/17, 0.017) & +8.3\% (12/17, 0.120) & +7.6\% (12/17, 0.031) & +9.0\% (9/17, 0.284) \\
Direct ridge & +28.0\% (17/17, 2e-5) & +15.3\% (17/17, 2e-5) & +23.2\% (15/17, 4e-4) & +14.2\% (13/17, 0.015) \\
Ridge + aug. & -1.1\% (8/17, 0.431) & +1.5\% (10/17, 0.306) & -1.1\% (8/17, 0.678) & +1.4\% (11/17, 0.487) \\
MLP & +39.3\% (17/17, 2e-5) & +14.7\% (13/17, 0.004) & +44.9\% (17/17, 2e-5) & +25.9\% (12/17, 0.031) \\
MLP + aug. & +33.9\% (17/17, 2e-5) & +10.0\% (12/17, 0.051) & +32.3\% (16/17, 2e-4) & +15.2\% (10/17, 0.353) \\
GRU & +38.7\% (17/17, 2e-5) & +15.7\% (15/17, 1e-4) & +36.8\% (15/17, 2e-4) & +15.7\% (12/17, 0.120) \\
GRU + aug. & +28.9\% (17/17, 2e-5) & +7.3\% (13/17, 0.017) & +26.1\% (15/17, 1e-4) & +8.5\% (10/17, 0.487) \\
Anchored scalar & +1.7\% (14/17, 0.006) & +0.2\% (9/17, 0.487) & +0.0\% (9/17, 0.611) & +0.4\% (10/17, 0.818) \\
Anchored free & +1.8\% (12/17, 0.011) & +0.7\% (10/17, 0.404) & +4.4\% (13/17, 0.013) & +1.9\% (10/17, 0.378) \\
Anchored free + aug. & +1.5\% (11/17, 0.020) & +0.2\% (10/17, 0.517) & +0.1\% (9/17, 0.782) & +0.8\% (9/17, 0.611) \\
 
\bottomrule
\end{tabular}
\end{table*}

\begin{table}[h]
\caption{Forecast error on held-out teams (nMSE; lower is better) over the first three windows (16--48\,s) and over all 16 (16--256\,s). Top pair: how often the most engaged pair is identified 128\,s ahead. Shaded: \model; bold: best in each column. Params: number of values fitted from the data. The anchored models use \model{}'s fitted anchors, which favours them. These errors pool all forecasts before averaging, whereas \cref{tab:baselines,tab:relative_detail} average per team first, as in \cref{tab:principles}; ratios of the values here therefore differ slightly from the percentages there.}
\label{tab:absolute}
\small
\setlength{\tabcolsep}{2pt}
\begin{tabular}{lrccccc}
\toprule
& & \multicolumn{2}{c}{Member and pair} & \multicolumn{2}{c}{Team} & Top\\
\cmidrule(lr){3-4}\cmidrule(lr){5-6}
Model & Params & 16--48\,s & 16--256\,s & 16--48\,s & 16--256\,s & pair\\
\midrule
\rowcolor{ours}ANCHOR (ours) & $52$ & 0.758 & \textbf{1.082} & 0.705 & \textbf{1.002} & 30\% \\
ANCHOR + observer & $56$ & 0.759 & \textbf{1.082} & 0.707 & 1.004 & 30\% \\
Attractor only & $21$ & 0.848 & 1.099 & 0.775 & 1.009 & 30\% \\
\midrule
Persistence & $0$ & 1.049 & 1.627 & 0.913 & 1.416 & \textbf{31\%} \\
Recent average & $0$ & 0.927 & 1.330 & 0.807 & 1.206 & \textbf{31\%} \\
Session mean & $0$ & 0.971 & 1.229 & 1.030 & 1.357 & 30\% \\
\midrule
VAR(2) & $820$ & 0.899 & 1.258 & 0.850 & 1.184 & 18\% \\
VAR(2) + aug. & $820$ & 0.826 & 1.242 & 0.780 & 1.170 & 15\% \\
Direct ridge & $43{,}840$ & 0.993 & 1.281 & 0.902 & 1.195 & 16\% \\
Ridge + aug. & $43{,}840$ & \textbf{0.754} & 1.110 & \textbf{0.703} & 1.044 & 27\% \\
MLP & $14{,}944$ & 1.084 & 1.291 & 1.073 & 1.340 & 17\% \\
MLP + aug. & $14{,}944$ & 1.057 & 1.258 & 1.004 & 1.253 & 18\% \\
 
\midrule
Anchored scalar & $28$ & 0.771 & 1.089 & 0.702 & 1.014 & --\\
Anchored free & $434$ & 0.773 & 1.096 & 0.735 & 1.030 & --\\
Anchored free + aug. & $434$ & 0.770 & 1.089 & 0.706 & 1.021 & --\\
GRU & $27{,}168$ & 1.066 & 1.283 & 0.992 & 1.213 & --\\
GRU + aug. & $27{,}168$ & 0.994 & 1.202 & 0.926 & 1.158 & --\\
 
\bottomrule
\end{tabular}
\end{table}

\begin{figure}[t]
\centering
\includegraphics[width=\columnwidth]{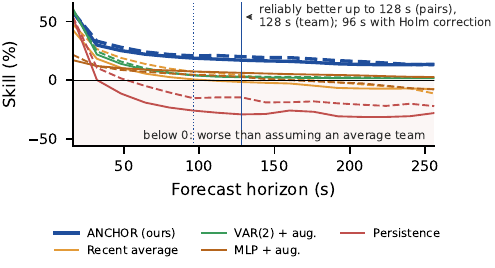}
\caption{How far ahead \model{} is useful. Skill is the error reduction relative to assuming an average training team (higher is better; solid: member and pair level, dashed: team level). Vertical lines mark the last horizon up to which \model{} beats that reference with $p<0.05$ at every horizon (solid: pairs, dashed: team, dotted: both levels with Holm correction for the 16 tests).}
\label{fig:skill}
\end{figure}

\paragraph{How far ahead.} \Cref{fig:skill} compares \model{} with assuming an average training team, by horizon. Long-range errors above 1, as in \cref{fig:horizon}, still carry skill. The error is scaled by the variation of the training teams, and a held-out team's later windows differ from the training average by more than that variation: at 128\,s, assuming an average team gives \hzRefOneTwoEight{} against \hzAnchorOneTwoEight{} for \model.

\begin{figure}[t]
\centering
\includegraphics[width=\columnwidth]{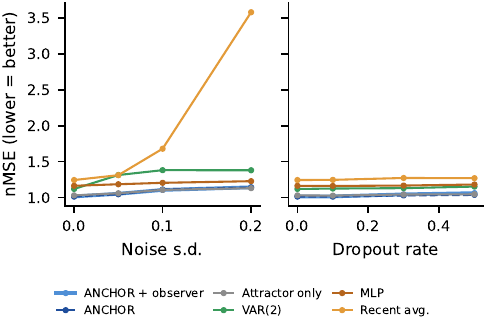}
\caption{Member and pair error under additive sensor noise (left) and per-modality window dropout (right) when training and test inputs are corrupted alike.}
\label{fig:robust}
\end{figure}

\paragraph{Sensor corruption.} We corrupt input windows, never targets, with Gaussian noise or with dropout of each modality of a window with probability $p$. When models are trained with the same corruption, \model{} or its observer has lower error than the learned baselines tested (\cref{fig:robust,tab:robust}), and the attractor-only control is within about 0.01 of it, slightly lower at the two strongest noise levels; augmented ridge was not part of this test. With noise of standard deviation 0.2 its nMSE rises from \rbCleanPh{} to \rbGaussTwoPh, against \rbGaussTwoVar{} for VAR. Without retraining, dropout remains benign, but strong noise hurts \model{} far more than VAR (\rbGaussTwoShiftPh{} against \rbGaussTwoShiftVar), because its resting point averages the corrupted windows, so noise must be matched in training.

\begin{figure}[t]
\centering
\includegraphics[width=\columnwidth]{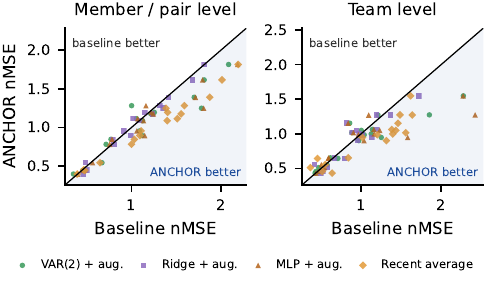}
\caption{Per-team error of \model{} against the strongest baselines (one point per held-out team and baseline). Points in the shaded area below the diagonal are teams for which \model{} is better.}
\label{fig:perteam}
\end{figure}

\paragraph{Per team and per quantity.} \Cref{fig:perteam} compares \model{} with the strongest baselines team by team, and \cref{tab:team} breaks the team-level error down by quantity; attention density is the hardest quantity for every model.

\paragraph{Top pair.} \model{} predicts \emph{how much} a pair will interact better than \emph{which} pair will lead, where persistence is level or slightly stronger (\topDyadPers\% against \topDyadBest\%).

\begin{figure}[h]
\centering
\includegraphics[width=\columnwidth]{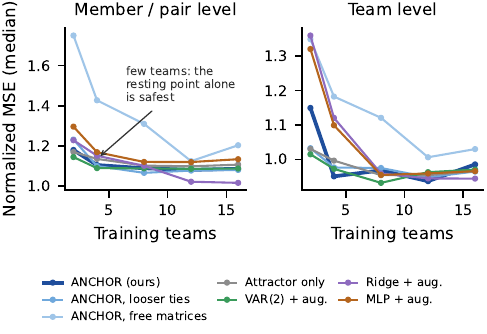}
\caption{Error versus the number of training teams, as the median over held-out teams and random training subsets. With very few teams the resting point alone is the safest forecast. \model, looser ties and \model, free matrices are the permutation-equivariant and unconstrained variants of \cref{app:model}.}
\label{fig:de}
\end{figure}

\paragraph{How much data the model needs.} \Cref{fig:de} trains every model on \deMinTrain--\deMaxTrain{} teams. On mean error \model{} stays below augmented ridge at every size, significantly up to eight training teams at member and pair level and up to four at team level; with more teams the two are on par, and at 12--16 teams the median member and pair error favours ridge. \model{} reaches the attractor-only control from about eight teams at member and pair level and about 12 at team level, without a significant gain there. The variant with free matrices needs the most data.

\paragraph{What the structure adds.} All anchored baselines take \model{}'s fitted anchors, so they share its resting point. A scalar relaxation fits one decay and one port gain per behaviour and per team coordinate by least squares. A free linear map fits $z_{h+1}=Az_h+b\bar u_h$ with ridge regression. The GRU has 32 hidden units, reads the last five windows, and adds the session mean and the task input to a linear output layer. Over 170 retrainings on 2, 4, 8 and 16 teams, $\rho$ reached at most \rtRhoMax{} for \model{} by construction, and the free map had a spectral radius of at most \rtFreeMax{} empirically. At small sizes the scalar relaxation had the lowest member and pair error, which shows that the fewest parameters win when data are scarcest.

\subsection{The fitted model and real teams}\label{app:teams}

This part gives the fitted values behind \cref{sec:teams} and further checks of how real teams move: raw energy, the energy balance on one forecast, the timing of the phase, richer switching and the scale of seconds.

\begin{table}[t]
\caption{Fitted model on all teams. Anchor weights; shift of the resting point per unit of remaining work (negative: it rises as images are decided); and recovery time constants in seconds in consolidation and exploration.}
\label{tab:structure}
\footnotesize
\setlength{\tabcolsep}{2.6pt}
\begin{tabular}{lcccccc}
\toprule
Coordinate & $w_{\mathrm{rec}}$ & $w_{\mathrm{ses}}$ & $w_{\mathrm{pop}}$ & Shift & Cons. & Expl.\\
\midrule
Speaking (member) & 0.44 & 0.19 & 0.37 & -0.058 & 26 & 37 \\
Attention (pair) & 0.18 & 0.81 & 0.01 & -0.073 & 17 & 28 \\
Proximity (pair) & 0.10 & 0.41 & 0.50 & -0.066 & 28 & 55 \\
Team speaking level & 0.34 & 0.13 & 0.53 & -0.042 & 38 & 32 \\
Team attention density & 0.35 & 0.58 & 0.07 & -0.060 & 22 & 43 \\
Team proximity density & 0.19 & 0.46 & 0.35 & -0.111 & 15 & 78 \\
Team speaking balance & 0.43 & 0.02 & 0.56 & -0.254 & 12 & 22 \\
 
\bottomrule
\end{tabular}
\end{table}

\paragraph{Raw energy on real teams.} The fading of \cref{fig:memory} appears across many windows, and noise hides it in any one of them. After the \enOrigins{} windows with the highest energy, the observed energy measured from the same resting point is \enObsThree{} times its starting value 48\,s later, against \enSurrThree{} for shuffled sessions and \enPredThree{} for the forecast, and at most lags real and shuffled sessions have similar energy. Energy is quadratic, so window-to-window noise always adds to it, whereas the forecast is the noise-free expected path. The signed projection of \cref{eq:memory}, which we chose after seeing this result, is the measure that matches an expected path. Noise in the starting window still lowers every $m_h$ by the same factor, because it inflates the denominator of \cref{eq:memory} but cancels in the numerator, so the levels of $m_h$ may be slightly understated while the ratio between successive steps, about 30\% removed per 16\,s, is unaffected. Energy is also distinct from task progress (rank correlation \rhoHProgress).

\begin{figure}[h]
\centering
\includegraphics[width=\columnwidth]{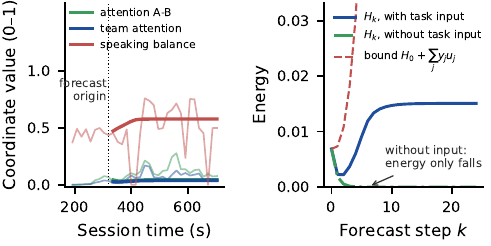}
\caption{The energy balance on one held-out forecast. Left: a 24-window forecast (thick) from the dotted origin, with the observed values (faint), for three coordinates on their 0--1 scale. Right: the forecast's energy $H_k$, measured from the resting point at the origin, with task input (solid) and for the same forecast without task input (dash-dotted), and the bound of \cref{prop:passivity}, the initial energy plus the energy supplied by task progress (dashed). Without input the energy only falls; with input every rise comes from the task and stays below the bound.}
\label{fig:rollout}
\end{figure}

\paragraph{The energy balance on one forecast.} \Cref{fig:rollout} follows one 24-window forecast for a held-out team to show what \cref{prop:passivity} says. The energy is measured from the resting point at the forecast origin. Without task input, the same forecast loses energy at every step and is back at rest within a few windows, which is the half of the proposition that makes the model safe to iterate: a forecast cannot drift away from rest on its own. With task input, the energy first falls in the same way and then rises and levels off, because the forecast change in remaining work supplies energy through the port and moves the forecast toward a shifted resting point (\cref{lem:shift}). The dashed bound, the initial energy plus the energy supplied by the task, is loose here, because it leaves out the energy that regulation removes; what it guarantees is that every rise in energy comes from the task.

\paragraph{When the phase changes.} Remaining work crosses the phase threshold once in \phOnce{} of \nGroups{} sessions, and for the first time at a median of \phFirstSec\,s (interquartile range \phFirstLow--\phFirstHigh\,s), in sessions of 5.6--19.7\,min. Of all forecasts, \phSoon\% meet a switch within 48\,s and \phAny\% within 256\,s. The forecast phase matches the phase later observed at \phAgree\% of horizons. The phase therefore works much like an early and late part of the session, and the slower recovery in exploration also reflects time in session. The task signal also partly reaches the forecast through the anchors: the median within-team correlation with remaining work is \phCorrWin{} for the attention density of a window and \phCorrRest{} for its resting point, and a session average also trends with time. In a scalar relaxation, the task term has about one eighth of the magnitude of the deviation for speaking, one fifth for proximity and balance, and one half for attention. These findings are consistent with phase and port being on par in accuracy.

\paragraph{Richer switching and nonlinear steps.} With \model{}'s anchors, we compared the scalar relaxation with 22 switched variants, each scored at four settings. They switched decay, and optionally shifted the resting point, by task phase, by behavioural clusters or by MURMR's interaction states~\cite{murmr}, or they changed decay with the sign or size of the deviation. All the switched variants remain \pH{} systems. After Holm correction, the scalar relaxation matched all of them, and the largest gain, 2.9\% (task phase with a shifted resting point, team level, 16--256\,s), was below significance. A generous bound confirms this: when three hidden states of team behaviour are inferred from the whole held-out session and known at every future step, with the resting point shifting by state, error falls by 3.6--6.0\% across levels and horizons, although this bound sees the future. MURMR's states, read with its released encoder, are stable for most pairs within a session, because several encoder inputs are pair-level summaries, so here they describe the pair more than a moment in time. A free nonlinear correction with no stability constraint grew without bound (error +51\% at 16--48\,s), which illustrates the value of built-in stability.

\begin{table*}[t]
\caption{Second-scale diagnostic on the 1\,s interaction table: time-irreversibility of the lagged cross-covariance (signature of lossless exchange $J$), share of rotational modes of a fitted linear model (signature of $J$ or inertia), and the most negative autocorrelation within 60\,s (signature of inertia), for real teams and for surrogate teams assembled from members or pairs of different teams. The proximity signal here is the continuous pair distance. $p$: one-sided Mann--Whitney test that real teams show more of the signature.}
\label{tab:app_timescale}
\small
\setlength{\tabcolsep}{4pt}
\begin{tabular}{lrcccccccc}
\toprule
& & \multicolumn{3}{c}{Irreversibility} & \multicolumn{3}{c}{Rotational share} & \multicolumn{2}{c}{Min.\ autocorrelation}\\
\cmidrule(lr){3-5}\cmidrule(lr){6-8}\cmidrule(lr){9-10}
Signal & Bin (s) & Team & Surr. & $p$ & Team & Surr. & $p$ & Team & Surr.\\
\midrule
Speaking & 1 & 0.279 & 0.293 & 0.715 & 0.010 & 0.014 & 0.782 & -0.063 & -0.081 \\
Speaking & 2 & 0.287 & 0.309 & 0.767 & 0.034 & 0.034 & 0.326 & -0.066 & -0.092 \\
Speaking & 4 & 0.270 & 0.288 & 0.727 & 0.093 & 0.087 & 0.220 & -0.069 & -0.101 \\
Speaking & 8 & 0.298 & 0.295 & 0.624 & 0.185 & 0.185 & 0.473 & -0.052 & -0.084 \\
Proximity & 1 & 0.127 & 0.135 & 0.480 & 0.021 & 0.016 & 0.063 & +0.030 & -0.038 \\
Proximity & 2 & 0.134 & 0.146 & 0.564 & 0.036 & 0.035 & 0.379 & +0.033 & -0.037 \\
Proximity & 4 & 0.131 & 0.140 & 0.566 & 0.061 & 0.072 & 0.875 & +0.039 & -0.039 \\
Proximity & 8 & 0.152 & 0.170 & 0.681 & 0.133 & 0.156 & 0.872 & +0.067 & -0.005 \\
Attention & 1 & 0.438 & 0.484 & 0.870 & 0.088 & 0.088 & 0.627 & -0.065 & -0.081 \\
Attention & 2 & 0.418 & 0.468 & 0.906 & 0.102 & 0.142 & 0.956 & -0.077 & -0.099 \\
Attention & 4 & 0.385 & 0.423 & 0.886 & 0.244 & 0.224 & 0.264 & -0.074 & -0.106 \\
Attention & 8 & 0.374 & 0.433 & 0.977 & 0.327 & 0.338 & 0.443 & -0.050 & -0.094 \\
 
\bottomrule
\end{tabular}
\end{table*}

\paragraph{Is there exchange or inertia at the scale of seconds?} The model finds no evidence of lossless routing or inertia at 16\,s, but such structure might still exist at finer scales. To test this, we analysed the per-second interaction table at bins of 1--8\,s (\cref{tab:app_timescale}) and compared real teams with surrogate teams assembled from unrelated members. For every signal and bin size, real teams show no more time-irreversibility, rotation or oscillation than the surrogates, and proximity and attention are, if anything, smoother and more reversible. The dissipative picture therefore holds down to one second for these two signals. Conversation could not be tested conclusively. In the per-second speaking signal, simultaneous speech occurs as often as if members spoke independently (median ratio of observed to independent overlap \overlapRatio), whereas genuine turn-taking would make overlap rarer than chance. Cross-talk between co-located headsets and one-second bins that span speaker changes are likely causes, so testing conversational exchange requires segment-level diarization free of cross-talk.

\subsection{Early warning}\label{app:warning}

\paragraph{Early warning, in full.} Warning times in \cref{tab:warning} largely reflect the \ewWindowSec\,s alarm window, since every model warns about as early.

\begin{figure*}[t]
\centering
\includegraphics[width=0.85\textwidth]{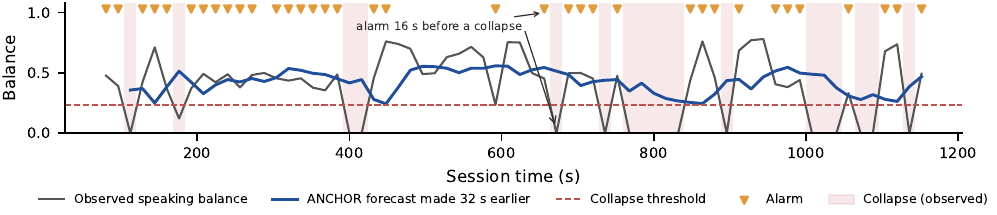}
\caption{Early warning on the longest session (team \caseTeam). Grey: observed speaking balance; blue: \model{} forecast made 32\,s earlier, which follows the slow drift of the balance rather than its window-to-window swings; dashed: collapse threshold; shaded: windows in collapse; triangles: alarms at the pooled \ewFar\% false-alarm level. Most collapses are preceded by an alarm, but this chronically uneven team also triggers many alarms that are not followed by a collapse.}
\label{fig:case}
\end{figure*}

\begin{table}[H]
\caption{Early-warning AUROC (0.5 = chance, 1 = perfect) for a breakdown at a single future window, by horizon: all origins / onsets only (breakdown absent at the origin).}
\label{tab:app_auroc}
\small
\setlength{\tabcolsep}{3pt}
\begin{tabular}{lcccccc}
\toprule
& \multicolumn{2}{c}{32\,s} & \multicolumn{2}{c}{64\,s} & \multicolumn{2}{c}{96\,s}\\
\cmidrule(lr){2-3}\cmidrule(lr){4-5}\cmidrule(lr){6-7}
Model & all & onset & all & onset & all & onset\\
\midrule
\emph{Conversation collapse} & & & & & & \\
\quad ANCHOR (ours) & 0.80 & 0.83 & 0.78 & 0.80 & 0.76 & 0.78 \\
\quad Attractor only & 0.81 & 0.83 & 0.78 & 0.79 & 0.76 & 0.78 \\
\quad Persistence & 0.76 & 0.73 & 0.75 & 0.72 & 0.72 & 0.69 \\
\quad Recent average & 0.80 & 0.82 & 0.78 & 0.78 & 0.76 & 0.78 \\
\quad VAR(2) + aug. & 0.76 & 0.74 & 0.78 & 0.78 & 0.79 & 0.80 \\
\quad Ridge + aug. & 0.78 & 0.77 & 0.75 & 0.74 & 0.75 & 0.75 \\
\quad MLP + aug. & 0.75 & 0.74 & 0.74 & 0.74 & 0.74 & 0.74 \\
\emph{Loss of shared focus} & & & & & & \\
\quad ANCHOR (ours) & 0.81 & 0.75 & 0.74 & 0.70 & 0.67 & 0.62 \\
\quad Attractor only & 0.80 & 0.74 & 0.74 & 0.70 & 0.68 & 0.61 \\
\quad Persistence & 0.82 & 0.73 & 0.74 & 0.70 & 0.73 & 0.66 \\
\quad Recent average & 0.81 & 0.74 & 0.76 & 0.72 & 0.72 & 0.64 \\
\quad VAR(2) + aug. & 0.72 & 0.67 & 0.63 & 0.59 & 0.59 & 0.55 \\
\quad Ridge + aug. & 0.73 & 0.68 & 0.72 & 0.68 & 0.68 & 0.59 \\
\quad MLP + aug. & 0.75 & 0.69 & 0.71 & 0.65 & 0.66 & 0.58 \\
\emph{Low proximity} & & & & & & \\
\quad ANCHOR (ours) & 0.71 & 0.67 & 0.61 & 0.63 & 0.58 & 0.61 \\
\quad Attractor only & 0.71 & 0.64 & 0.61 & 0.60 & 0.57 & 0.58 \\
\quad Persistence & 0.74 & 0.68 & 0.62 & 0.65 & 0.57 & 0.59 \\
\quad Recent average & 0.71 & 0.64 & 0.61 & 0.60 & 0.57 & 0.59 \\
\quad VAR(2) + aug. & 0.69 & 0.66 & 0.54 & 0.55 & 0.44 & 0.42 \\
\quad Ridge + aug. & 0.72 & 0.66 & 0.68 & 0.70 & 0.65 & 0.66 \\
\quad MLP + aug. & 0.67 & 0.60 & 0.60 & 0.58 & 0.59 & 0.58 \\
 
\bottomrule
\end{tabular}
\end{table}

\paragraph{Early warning by horizon.} \Cref{tab:app_auroc} complements \cref{tab:warning} for a single future window. For conversation collapse the leading models are close, and the augmented VAR is strongest at 96\,s. Augmented ridge leads for low proximity from 64\,s on, and the augmented VAR is weakest for loss of shared focus.

\begin{table}[h]
\caption{Sensitivity to the breakdown threshold (training percentile). Early warning: AUROC of \model{} / best other model for conversation collapse, loss of shared focus and low proximity. Intervention: settings (of \sensSettings) in which \model-predictive is the most cue-efficient policy, and its breakdown time removed per cue relative to periodic cueing at $\beta=0.5$\,s.d. The remaining settings are won by reactive cueing, and at the 35th percentile also by the forecast policy.}
\label{tab:sensitivity}
\scriptsize
\setlength{\tabcolsep}{3pt}
\begin{tabular}{lccccc}
\toprule
Threshold & Collapse & Focus & Low prox. & Predictive best & vs.\ periodic\\
\midrule
15th & 0.75 / 0.76 & 0.77 / 0.77 & 0.65 / 0.69 & 5/12 & 3.9$\times$ \\
25th (main) & 0.85 / 0.85 & 0.76 / 0.76 & 0.66 / 0.68 & 8/12 & 2.2$\times$ \\
35th & 0.75 / 0.77 & 0.77 / 0.79 & 0.66 / 0.69 & 8/12 & 1.4$\times$ \\
 
\bottomrule
\end{tabular}
\end{table}

\paragraph{Sensitivity to the breakdown threshold.} \Cref{tab:sensitivity} repeats early warning and the simulated interventions with breakdowns defined at the 15th and 35th training percentile. At every threshold, \model{} stays within 0.02 of the best model for conversation collapse and loss of shared focus and ahead of the learned baselines, and augmented ridge leads for low proximity. Collapse AUROC is 0.85 at the main threshold and about 0.75 at the other two. The advantage of \model-predictive over periodic cueing shrinks as breakdowns become more frequent, because a fixed schedule then hits a breakdown more often by chance.

\subsection{Simulated cues}\label{app:cues}

\paragraph{Simulation protocol.} Each simulated episode starts from the first six windows of a held-out team and runs for 40 windows (about 11\,min). At every window the plant, either \model{} fitted on all teams or an independently learned VAR(2) on relabel-augmented data, predicts the next window, and an error drawn at random from the plant's one-step errors on real data is added, so the simulated team varies as real teams do. A cue adds $\beta\in\{0.25,0.5,1\}$ training standard deviations to its targets in the window it is issued: speaking balance for balance cues, and attention density with all six pair attentions for attention cues. The forecast and \model-predictive policies use \model{} fitted without the simulated team, and their trigger levels are set so that, on the training teams, they would fire about as often as periodic cueing; in simulation their rates differ from periodic cueing (\cref{tab:intervention}), which is why we report the effect per cue. Each team is simulated with 30 random seeds, and every policy sees the same random draws. We report the share of the episode in breakdown and the breakdown time removed per cue relative to no cueing.

\paragraph{Interventions for every effect size.} \Cref{tab:app_intervention} gives the full simulation results behind \cref{fig:intervention}. Larger cue effects raise every policy's efficiency. \model-predictive is the most efficient policy in \ivMpcWins{} of the 12 settings, and reactive cueing in the other four: attention cues on the VAR plant at every effect size and balance cues on the VAR plant at $\beta=1$.

\begin{table}[t]
\caption{Time in breakdown and cue budget at cue effect $\beta=0.5$\,s.d. Each cell gives the share of the episode spent in the breakdown (\%, lower is better), with the number of cues per 10\,min in parentheses. Shaded: \model-predictive, the most selective policy.}
\label{tab:intervention}
\scriptsize
\setlength{\tabcolsep}{2.2pt}
\begin{tabular}{llccccO}
\toprule
Cue & Plant & None & Periodic & Reactive & Forecast & \model-predictive\\
\midrule
Balance & ANCHOR & 13.3 & 11.4 (4.7) & 11.4 (3.8) & 10.8 (5.1) & 12.1 (1.8) \\
Attention & ANCHOR & 11.5 & 5.7 (4.7) & 5.6 (2.3) & 5.3 (1.9) & 5.6 (1.7) \\
Balance & VAR & 12.6 & 11.0 (4.7) & 10.4 (3.4) & 10.0 (4.8) & 11.5 (1.7) \\
Attention & VAR & 12.5 & 7.6 (4.7) & 7.0 (2.8) & 9.0 (1.9) & 9.7 (1.5) \\
 
\bottomrule
\end{tabular}
\end{table}

\begin{table*}[t]
\caption{Simulated interventions for every assumed cue effect $\beta$ (training s.d.): share of the episode in the breakdown (\%, lower is better) / breakdown time removed per cue (percentage points, higher is better). \model-predictive is the most efficient policy in \ivMpcWins{} of the 12 settings, and reactive cueing in the others.}
\label{tab:app_intervention}
\small
\begin{tabular}{lllccccc}
\toprule
Cue & Plant & $\beta$ & None & Periodic & Reactive & Forecast & \model-predictive\\
\midrule
Balance & ANCHOR & 0.25 & 13.3 & 12.4 / 0.18 & 12.3 / 0.24 & 12.0 / 0.23 & 12.8 / 0.34 \\
Balance & ANCHOR & 0.50 & 13.3 & 11.4 / 0.38 & 11.4 / 0.47 & 10.8 / 0.47 & 12.1 / 0.62 \\
Balance & ANCHOR & 1.00 & 13.3 & 9.9 / 0.68 & 9.8 / 0.94 & 8.9 / 0.93 & 9.8 / 1.06 \\
Attention & ANCHOR & 0.25 & 11.5 & 8.2 / 0.65 & 7.5 / 1.37 & 7.2 / 1.62 & 7.2 / 1.64 \\
Attention & ANCHOR & 0.50 & 11.5 & 5.7 / 1.15 & 5.6 / 2.41 & 5.3 / 2.98 & 5.6 / 3.22 \\
Attention & ANCHOR & 1.00 & 11.5 & 3.1 / 1.68 & 4.0 / 3.81 & 3.9 / 5.08 & 4.2 / 5.28 \\
Balance & VAR & 0.25 & 12.6 & 11.8 / 0.16 & 11.4 / 0.30 & 11.1 / 0.27 & 12.1 / 0.33 \\
Balance & VAR & 0.50 & 12.6 & 11.0 / 0.31 & 10.4 / 0.58 & 10.0 / 0.50 & 11.5 / 0.60 \\
Balance & VAR & 1.00 & 12.6 & 9.7 / 0.56 & 8.9 / 1.06 & 8.4 / 0.93 & 9.2 / 1.04 \\
Attention & VAR & 0.25 & 12.5 & 9.4 / 0.62 & 8.5 / 1.18 & 9.7 / 1.16 & 9.8 / 1.15 \\
Attention & VAR & 0.50 & 12.5 & 7.6 / 0.97 & 7.0 / 1.81 & 9.0 / 1.73 & 9.7 / 1.69 \\
Attention & VAR & 1.00 & 12.5 & 5.2 / 1.45 & 5.8 / 2.49 & 8.6 / 2.41 & 9.6 / 2.26 \\
 
\bottomrule
\end{tabular}
\end{table*}

% % -----------------------------------------------------------------------------
% \section{Reproducibility}\label{app:repro}
% % -----------------------------------------------------------------------------

% \begin{sloppypar}
% \texttt{run\_full\_pipeline.sh} reproduces all results from the MURMR dataset. It runs \texttt{ph\_team\_experiments.py} for the model families, baselines, data efficiency, robustness and fitted structure; \texttt{add\_observer\_rows.py}, \texttt{check\_sym\_pop.py}, \texttt{deviation\_memory.py} and \texttt{energy\_on\_real\_teams.py} for the additional checks; \texttt{ph\_interventions.py} at three thresholds for early warning and simulated cues; \texttt{closed\_loop\_bound.py}; \texttt{ph\_timescale\_diagnostic.py}; \texttt{deviation\_by\_behaviour.py} for the forecasts by behaviour in \cref{sec:real}; \texttt{structure\_tests.py} for the anchored baselines, retrainings, autoregressive control and resting-point floor; and \texttt{phase\_timing.py}, \texttt{phase\_nonlinear\_tests.py} and \texttt{phase\_inferred\_tests.py} for the phase results of \cref{app:teams}. \texttt{make\_paper\_assets.py}, \texttt{make\_anchor\_figs.py} and \texttt{make\_structure\_assets.py} then generate the figures, tables and macros. On one laptop thread a 16-step forecast takes \timeForecast\,ms and fitting on 16 teams takes \timeFit\,s, well within the 16\,s stride of the live loop.
% \end{sloppypar}

\end{document}